\documentclass{article}

\usepackage[english]{babel}
\usepackage{amsmath}
\usepackage[normalem]{ulem}
\usepackage{amssymb}
\usepackage{color}
\usepackage[font=small]{caption}
\usepackage[font=small]{subcaption}
\usepackage[makeroom]{cancel}
\usepackage{tikz}
\usepackage[compat=1.1.0]{tikz-feynman}
\usepackage{mathptmx}
\usetikzlibrary{arrows}
\usepackage{float}
\usepackage{inputenc}
\usepackage[T1]{fontenc}
\usepackage{lmodern}
\usepackage{geometry}

\usepackage{microtype}
\usepackage{booktabs}
\usepackage{array}
\usepackage{tabularx}
\newcommand{\formfactortablestyle}{%
  \small
  \renewcommand{\arraystretch}{1.15}%
  \setlength{\tabcolsep}{5pt}%
}

\usepackage{jheppub}

\usepackage{graphicx} 
\usepackage{dcolumn}
\usepackage[dvipsnames,table]{xcolor}   

\usepackage{mathrsfs}                   

\usepackage{slashed}

\usepackage[detect-all]{siunitx}
\title{Probing Heavier-Baryon Oscillations through Light-Baryon Mixing}

\author[a,b]{Arnau Bas i Beneito,}
\affiliation[a]{Departament de F\'isica Te\`orica, Universitat de Val\`encia, 46100 Burjassot, Spain}
\affiliation[b]{Institut de F\'isica Corpuscular (CSIC-Universitat de Val\`encia), Parc Cient\'ific UV, C/Catedr\'atico Jos\'e Beltr\'an, 2, E-46980 Paterna, Spain}
\emailAdd{arnau.bas@ific.uv.es}

\author[c,d]{Svjetlana Fajfer,}
\affiliation[c]{Department of Physics, University of Ljubljana, Jadranska 19, 1000 Ljubljana, Slovenia}
\affiliation[d]{Jo\v zef Stefan Institute, Jamova 39, 1000 Ljubljana, Slovenia}
\emailAdd{svjetlana.fajfer@ijs.si}

\author[e]{Alexey A. Petrov}
\affiliation[e]{Department of Physics and Astronomy, University of South Carolina, Columbia, South Carolina 29208, USA}
\emailAdd{apetrov@sc.edu}

\abstract{
Motivated by the existing direct limits on $\Xi_b^0$ and $\Xi_c^0$ oscillations, we investigate how Standard Model weak interactions connect baryon-number-violating mixing in heavy neutral baryons to the much more strongly constrained neutron-antineutron and $\Lambda-\bar\Lambda$ mixing. We focus on $\Xi_b^0$, $\Lambda_b^0$, $\Xi_c^0$, and $\Omega_c^0$ oscillations and derive indirect constraints on their mass-mixing parameters from existing limits on $n-\bar n$ oscillations and on $\Lambda-\bar\Lambda$ mixing, the latter inferred from dinucleon-decay searches. The weak baryon-to-baryon transition amplitudes are estimated by saturating the relevant hadronic correlators with low-lying intermediate baryons. The required weak vertices are modelled using semileptonic form factors from lattice QCD, quark models, and QCD sum rules. Since these form factors must be continued outside their physical semileptonic domains and the resulting loop integrals depend on a UV cutoff, our results are order-of-magnitude estimates subject to sizeable hadronic uncertainties. For a benchmark cutoff of $10~{\rm GeV}$, the indirect bounds on the $\Xi_b^0$ and $\Xi_c^0$ mixing parameters are roughly three and five orders of magnitude stronger than the direct LHCb and Belle limits, indicating that light-baryon $|\Delta B|=2$ searches can probe heavy-baryon oscillations far beyond present direct sensitivities.}

\begin{document}

\maketitle

\section{Introduction}
\label{sec:bnv-introduction}

Observing baryon-number violation by two units would provide direct evidence for interactions beyond the Standard Model and an important clue to the origin of the matter--antimatter asymmetry of the Universe~\cite{Sakharov:1967dj}. Neutral baryon--antibaryon oscillations offer a probe of such interactions complementary to searches for ordinary proton decay~\cite{Weinberg:1979sa,Wilczek:1979hc,Abbott:1980zj,Arnold:2012sd}. Although neutron--antineutron oscillations provide an exceptionally sensitive test~\cite{Kuzmin:1970nx,Kuo:1980ew,Mohapatra:1980qe,Caswell:1982qs,Rao:1982gt,Rao:1983sd,Phillips:2014fgb}, they directly constrain only particular combinations of light-quark operators. Interactions involving strange, charm, or bottom quarks can have different flavour structures, motivating complementary searches in other neutral-baryon systems. Several scenarios have been discussed in which neutron oscillations and baryon-number-violating nuclear decays are suppressed by flavour symmetries or hierarchical couplings, while considerably larger mixings involving strange or heavy quarks remain possible~\cite{Kuzmin:1996sy,Nikolidakis:2007fc,Baldes:2011mh,Smith:2011rp,Durieux:2012gj,McKeen:2015cuz,Aitken:2017wie,Bittar:2024nrn,Beneito:2025ond,Beneito:2025fzf,BasiBeneito:2026feb}. Such possibilities have also been connected to CP violation and baryogenesis~\cite{McKeen:2015cuz,Aitken:2017wie}, providing additional motivation for direct searches.

The best-studied example is neutron--antineutron oscillation~\cite{Baldo-Ceolin:1994hzw,Super-Kamiokande:2020bov}, while searches have also been performed for $\Lambda-\bar\Lambda$ oscillations~\cite{BESIII:2024gcd} and for heavy neutral baryons. The LHCb Collaboration performed the first search for baryon-number-violating $\Xi_b^0-\bar{\Xi}_b^0$ oscillations, using the decay $\Xi_b^0\to\Xi_c^+\pi^-$ and its charge-conjugate final state to identify the baryon number at decay~\cite{LHCb:2017vth}. More recently, the Belle Collaboration searched for the baryon-number-violating decay $B^- \to \bar{\Xi}_c^0\bar{\Lambda}_c^-$ and for $\Xi_c^0-\bar{\Xi}_c^0$ oscillations, finding no signal and setting upper limits on both processes~\cite{Belle:2024cab}. The corresponding direct experimental limits on heavy-baryon oscillations are summarised in Table~\ref{tab:heavy_baryon_osc}. Their interpretation as mixing limits assumes negligible width splitting and that direct baryon-number-violating production or decay does not supply the wrong-sign signal.

\begin{table}[!htbp]
\centering
\formfactortablestyle
\begin{tabular}{c c c c}
\toprule
Baryon
& Experiment
& $|\delta m_{\mathcal B}|$ [GeV]
& $\tau_{\mathcal B-\bar {\mathcal B}}$ [s] \\
\midrule

$\Xi_b^0$
& LHCb~\cite{LHCb:2017vth}
& $5.1\times10^{-14}$
& $1.3\times10^{-11}$ \\

$\Xi_c^0$
& Belle~\cite{Belle:2024cab}
& $5.1\times10^{-13}$
& $1.3\times10^{-12}$ \\

\bottomrule
\end{tabular}
\caption{Current direct experimental limits at $95\%$ C.L. on heavy-baryon--antibaryon oscillations, expressed as upper bounds on the mass-mixing parameter $|\delta m_{\mathcal B} |$ and corresponding lower bounds on the oscillation time $\tau_{\mathcal B-\bar {\mathcal B}}$.}
\label{tab:heavy_baryon_osc}
\end{table}

Standard Model weak interactions connect baryons with different flavour content, so that a mass insertion $\delta m_{\mathcal B}$ for a heavy baryon $\mathcal B$ can induce an effective $n-\bar n$ or $\Lambda-\bar\Lambda$ transition. The experimental sensitivity to light-baryon oscillations is many orders of magnitude stronger than the direct sensitivity to heavy-baryon oscillations, motivating an assessment of the resulting indirect sensitivity. In particular, the Super-Kamiokande search for intranuclear $n-\bar n$ transitions implies, using the adopted nuclear suppression factor, a free-neutron oscillation time above $4.7\times10^8\,\mathrm{s}$ and a mass insertion below $1.4\times10^{-33}\,\mathrm{GeV}$~\cite{Super-Kamiokande:2020bov}. Dinucleon decay provides a further route to constraining strange-baryon mixing, subject to hadronic and nuclear uncertainties~\cite{Beneito:2025ond,Aitken:2017wie,Super-Kamiokande:2014hie,Super-Kamiokande:2015jbb}.

The purpose of this work is to quantify this connection for the four heavy baryons $\Lambda_b^0$, $\Xi_b^0$, $\Xi_c^0$, and $\Omega_c^0$. We consider the weak transitions $\Lambda_b^0\leftrightarrow n$ and $\Xi_b^0,\Xi_c^0,\Omega_c^0\leftrightarrow\Lambda$, using existing constraints on $|\delta m_n|$ and $|\delta m_\Lambda|$ to obtain indirect sensitivity estimates for their heavy-baryon counterparts. For $\Xi_b^0$ and $\Xi_c^0$, these estimates can be compared directly with the existing LHCb and Belle limits, while for $\Lambda_b^0$ and $\Omega_c^0$ they provide reference values for future searches. We focus on experimentally established, weakly decaying baryons, leaving extremely short-lived states and as-yet unobserved heavy baryons outside the scope of this work. The light $\Xi^0$ system is considered separately in Appendix~\ref{sec:Xiosc}.

The main challenge is hadronic, since the required off-diagonal weak baryon matrix elements are not directly known. We therefore employ an intermediate-state saturation approximation, with the weak vertices parametrised in terms of semileptonic form factors available in the literature. This allows us to estimate the heavy-to-light weak transition amplitudes entering the induced light-baryon mixing. The calculation extends the hadronic weak-transition mechanism of Ref.~\cite{Beneito:2025ond} to neutral heavy baryons and supplements the exploratory estimates of Ref.~\cite{Aitken:2017wie} with transition-specific form-factor input. The continuation of on-shell form factors to the kinematics of an off-diagonal self-energy, the truncation of intermediate states, and the regulator dependence introduce model dependence. The resulting sensitivities should therefore be understood as conditional hadronic estimates. Direct searches remain complementary because they probe heavy-baryon mixing without relying on these heavy-to-light transition estimates.

The framework is presented in Sec.~\ref{sec:framework}, while the numerical results and conclusions are given in Secs.~\ref{sec:numerics} and~\ref{sec:summary}, respectively. To keep the main discussion transparent, the definitions, parametrisations, conventions, and numerical input for the form factors are collected in Appendix~\ref{app:formfactors}. For completeness, in Appendix~\ref{sec:Xiosc} we also derive an indirect bound on $\Xi^0 - \bar \Xi^0$ oscillations. Since both $\Xi^0$ and $\Lambda$ belong to the light baryon octet, the corresponding weak transition can be described directly within baryon chiral perturbation theory (B$\chi$PT), without the intermediate-state approximation.

\section{Framework} \label{sec:framework}

Our approach builds on the hadronic weak-transition mechanism studied in Ref.~\cite{Beneito:2025ond}, which we extend here to the neutral heavy baryons $\Lambda_b^0$, $\Xi_b^0$, $\Xi_c^0$, and $\Omega_c^0$. We first establish the relation between heavy- and light-baryon mixing parameters and then describe the hadronic approximation employed to estimate the corresponding weak-transition amplitudes.

\subsection{Phenomenology of baryon-antibaryon mixing}
\label{sec:pheno}

For a neutral baryon $\mathcal B$, we define the $|\Delta B|=2$ mass insertion as

\begin{equation}
    {\cal L}_{|\Delta B|=2} = -\frac{1}{2} \left[ \delta m_\mathcal{B} \mathcal{B}^T C \mathcal{B} + \delta m_\mathcal{B}^* \mathcal{\bar B} C \mathcal{\bar B}^T \right]\, ,
\end{equation}
where $\delta m_{\mathcal B}$ is, in general, complex. Note that the phase of $\delta m_{\mathcal B}$ can enter a baryogenesis observable only through a rephasing-invariant relative phase between interfering amplitudes.~\cite{McKeen:2015cuz,Aitken:2017wie}. The short lifetimes of heavy baryons and the possibility of baryon-number-violating interactions with light degrees of freedom necessitate treating oscillation and decay together. In particular, baryon-number-violating interactions can generate both a mass difference and a lifetime difference between the propagating eigenstates. We formulate this problem using the same two-state language as for muonium-antimuonium conversion with a nonzero width difference~\cite{Conlin:2020veq,Petrov:2022wau} and for neutral-fermion mixing~\cite{McKeen:2015cuz}. Throughout, $\mathcal{B}$ denotes a neutral baryon, while $\Delta B$ denotes a change in baryon number. 

For a fixed spin projection in vacuum, let $|\psi(t)\rangle=a(t)|\mathcal{B}\rangle+b(t)|\mathcal{\bar B}\rangle$. The undecayed component evolves according to
\begin{equation}
    i\frac{d}{dt}\begin{pmatrix}a\\b\end{pmatrix} =\mathsf H_\mathcal{B} \begin{pmatrix}a\\b\end{pmatrix},\qquad \mathsf H_\mathcal{B}=\mathsf M_\mathcal{B}-\frac{i}{2}\boldsymbol\Gamma_\mathcal{B} =\begin{pmatrix} m_\mathcal{B}-\frac{i}{2}\Gamma_\mathcal{B} & M_{12}^\mathcal{B}-\frac{i}{2}\Gamma_{12}^\mathcal{B}\\ M_{12}^{\mathcal{B}*}-\frac{i}{2}\Gamma_{12}^{\mathcal{B}*} & m_\mathcal{B}-\frac{i}{2}\Gamma_\mathcal{B}
 \end{pmatrix}.
 \label{eq:intro-hamiltonian}
\end{equation}
Here CPT invariance fixes the equality of the diagonal entries, while $\mathsf M_\mathcal{B}$ and $\boldsymbol\Gamma_\mathcal{B}$ are Hermitian independently of whether CP is conserved. External magnetic fields and matter effects can lift the $\mathcal B-\bar{\mathcal B}$ degeneracy, and Eq.~\eqref{eq:intro-hamiltonian} assumes that their effect on the evolution is negligible. The two eigenstates and eigenvalues are
\begin{equation}
    |\mathcal{B}_{1,2}\rangle=p_\mathcal{B}|\mathcal{B}\rangle\pm q_\mathcal{B}|\bar {\mathcal B}\rangle,\qquad \lambda_{1,2}=m_{1,2}-\frac{i}{2}\Gamma_{1,2},\qquad |p_\mathcal{B}|^2+|q_\mathcal{B}|^2=1,
 \label{eq:intro-eigenstates}
\end{equation}
where the signs and square-root branch are chosen consistently, so that
\begin{align}
    \left(\frac{q_\mathcal{B}}{p_\mathcal{B}} \right)^2 &=\frac{M_{12}^{\mathcal{B}*}-i\Gamma_{12}^{\mathcal{B}*}/2}{M_{12}^\mathcal{B} -i \Gamma_{12}^\mathcal{B}/2 },\nonumber\\ \Delta m_\mathcal{B}-\frac{i}{2}\Delta \Gamma_\mathcal{B} &=2\sqrt{\left( M_{12}^\mathcal{B} -\frac{i}{2} \Gamma_{12}^\mathcal{B} \right)\left( M_{12}^{ \mathcal{B}*}-\frac{i}{2} \Gamma_{12}^{ \mathcal{B}*}\right)} \,.
 \label{eq:intro-diagonalization}
\end{align}
We define
\begin{equation}
    \Delta m_\mathcal{B}=m_1-m_2 \,,\quad \Delta\Gamma_\mathcal{B} =\Gamma_1-\Gamma_2 \, ,\quad \Gamma_\mathcal{B}=\frac{ \Gamma_1+\Gamma_2}{2} \, ,\quad x_\mathcal{B}=\frac{\Delta m_\mathcal{B}}{\Gamma_\mathcal{B}} \, ,\quad y_\mathcal{B}=\frac{ \Delta \Gamma_\mathcal{B}}{2\Gamma_\mathcal{B}} \, .
 \label{eq:intro-xy}
\end{equation}

Thus $\Gamma_\mathcal{B}$ is an average \emph{width}, and the two lifetimes are $\tau_{1,2} =1/ [\Gamma_\mathcal{B}(1\pm y_\mathcal{B})]$. A nonzero $y_\mathcal{B}$ is compatible with equal diagonal baryon and antibaryon widths. Our definition of $\Delta\Gamma_\mathcal{B}$ is explicit, so that reversing it reverses $y_\mathcal{B}$ without changing the conversion probabilities below.

Squaring Eq.~\eqref{eq:intro-diagonalization} gives the exact relations
\begin{equation}
    x_\mathcal{B}^2-y_\mathcal{B}^2 =\frac{4|M_{12}^\mathcal{B}|^2- | \Gamma_{12}^\mathcal{B} |^2}{\Gamma_\mathcal{B}^2} \, , \qquad x_\mathcal{B} y_\mathcal{B} =\frac{2\operatorname{Re}(M_{12}^\mathcal{B} \Gamma_{12}^{\mathcal{B}*})}{\Gamma_\mathcal{B}^2}\, .
 \label{eq:intro-exact-xy}
\end{equation}

If CP violation in mixing is neglected, the phases of the basis states may be chosen so that $M_{12}^\mathcal{B}$ and $\Gamma_{12}^\mathcal{B}$ are real and $q_\mathcal{B}/p_\mathcal{B}=1$. With the corresponding eigenstate labelling,
\begin{equation}
    x_\mathcal{B}=\frac{2M_{12}^\mathcal{B}}{\Gamma_\mathcal{B}}\, ,\qquad y_\mathcal{B}= \frac{\Gamma_{12}^\mathcal{B}}{\Gamma_\mathcal{B}} \, .
 \label{eq:intro-cp-limit}
\end{equation}

In the dispersive, single-insertion limit of the present analysis, $ |M_{12}^\mathcal{B}| =|\delta m_{\mathcal B}|$. Consequently, $|\Delta m_\mathcal{B}|=2|\delta m_{\mathcal B}|$ when $\Gamma_{12}^\mathcal{B}=0$: the off-diagonal mass insertion must be distinguished from the eigenstate mass splitting.

The physical origins of $x_\mathcal{B}$ and $y_\mathcal{B}$ are different. At energies below heavy new-particle thresholds, local six-quark operators can be written as
\begin{equation}
    \mathcal H_{|\Delta B|=2}^{\mathrm{loc}} =\sum_i\frac{C_i(\mu)}{\Lambda_{ \mathrm{BNV}}^5}\mathcal O_i^{(9)}(\mu)+\mathrm{h.c.}\, ,\qquad M_{12}^{\mathcal{B},\mathrm{loc}} =\frac{\langle \mathcal B|\mathcal H_{|\Delta B|=2}^{ \mathrm{loc} } | \bar{\mathcal B} \rangle}{2m_\mathcal{B}}\, .
 \label{eq:intro-local}
\end{equation}

The matrix element uses relativistically normalised states, and $ \Lambda_{ \mathrm{BNV}}$ is distinct from the regulator used in the hadronic calculation. Dispersive contributions from nonlocal insertions must be added when present. The absorptive part instead follows from the on-shell cut of the transition amplitude. With $A_f=\mathcal A(\mathcal B\to f)$ and $\bar A_f=\mathcal A(\bar {\mathcal B}\to f)$, unitarity gives
\begin{equation}
    \Gamma_{12}^\mathcal{B}=\frac{1}{2m_\mathcal{B}}\sum_f\int d\Phi_f\,A_f^*\bar A_f, \qquad \Gamma_\mathcal{B}= \frac{1}{2m_\mathcal{B}}\sum_f\int d\Phi_f\,|A_f|^2,
\label{eq:intro-optical}
\end{equation}
where $d\Phi_f$ includes energy--momentum conservation and the sum includes final-state spins. The amplitudes are the irreducible decay amplitudes, with propagation through the $\mathcal{B} -\bar { \mathcal{B}}$ mixing system excluded. Thus only common final states contribute to $\Gamma_{12}^\mathcal{B}$. Baryon-number-conserving weak decays alone produce disjoint baryon and antibaryon final states and do not generate this overlap. It can arise from interference between a baryon-number-conserving amplitude and a $|\Delta B|=2$ decay amplitude, from two $|\Delta B|=1$ amplitudes where such interactions exist, or from decays to shared light new states. A complex Wilson coefficient in a local mass term is not, by itself, an absorptive contribution. A prediction for $y_\mathcal{B}$ therefore requires decay information in addition to the mass insertion. Positivity of the decay matrix implies $|\Gamma_{12}^\mathcal{B}| \leq\Gamma_\mathcal{B}$.

The time evolution of an initially pure baryon is
\begin{equation}
    |\mathcal B(t)\rangle=g_+(t)|\mathcal B\rangle+ \frac{q_\mathcal{B}}{p_\mathcal{B}}g_-(t)| \bar{ \mathcal B}\rangle,\qquad g_\pm(t)=\frac{e^{-i\lambda_1t}\pm e^{-i\lambda_2t}}{2} \, ,
 \label{eq:intro-time}
\end{equation}
with $p_\mathcal{B}/q_\mathcal{B}$ replacing $q_\mathcal{B}/p_\mathcal{B}$ for an initially pure antibaryon. Direct evaluation yields
\begin{equation}
    |g_\pm(t)|^2=\frac{e^{-\Gamma_\mathcal{B}t}}{2} \left[ \cosh(y_\mathcal{B} \Gamma_\mathcal{B} t)\pm \cos(x_\mathcal{B} \Gamma_\mathcal{B}t)\right] \, .
 \label{eq:intro-gsquared}
\end{equation}
In particular, the probability that the state is an \emph{undecayed} antibaryon at time $t$ is
\begin{align}
    P_{\mathcal B\to \bar{\mathcal B}}(t) &=\left|\frac{q_\mathcal{B}}{p_\mathcal{B}}\right|^2 \frac{e^{-\Gamma_\mathcal{B}t}}{2} \left[ \cosh(y_\mathcal{B} \Gamma_\mathcal{B} t)-\cos(x_\mathcal{B}\Gamma_\mathcal{B} t)\right]\nonumber\\
    &=\left|\frac{q_\mathcal{B}}{p_\mathcal{B}} \right|^2 e^{-\Gamma_\mathcal{B} t} \frac{x_\mathcal{B}^2+y_\mathcal{B}^2}{4}(\Gamma_\mathcal{B}t)^2 +\mathcal O\!\left(e^{-\Gamma_\mathcal{B}t}(x_\mathcal{B}^4 +y_\mathcal{B}^4)(\Gamma_\mathcal{B}t )^4\right) \, .
 \label{eq:intro-probability}
\end{align}
The expansion requires $|x_\mathcal{B}|\Gamma_\mathcal{B}t,|y_\mathcal{B}|\Gamma_\mathcal{B}t\ll1$. For $y_\mathcal{B}=0$ and $|q_\mathcal{B}/ p_\mathcal{B}|=1$, this becomes $e^{-\Gamma_\mathcal{B}t}\sin^2(|\delta m_{\mathcal B}|t)$, recovering the conventional oscillation time $\tau_{\mathcal B -\bar{\mathcal B}}=1/|\delta m_{\mathcal B}|$ (or $\hbar/|\delta m_{\mathcal B}|$ in seconds). This time scale is distinct from the decay lifetime $\tau_{\mathcal B}=1/ \Gamma_\mathcal{B}$.

To connect this result to a measured wrong-sign decay yield, consider a baryon-specific tag $f$ and its conjugate $\bar f$, with no direct wrong-sign decay amplitudes and equal right-sign partial widths $\Gamma_{\mathrm{tag}}$. Then $dP_{\mathrm{WS}}/dt =\Gamma_{\mathrm{tag}} |q_\mathcal{B}/p_\mathcal{B}|^2 |g_-|^2$ and $dP_{\mathrm{RS}}/dt=\Gamma_{\mathrm{tag}}|g_+|^2$. For uniform acceptance, integration gives
\begin{equation}
    I_\pm\equiv\int_0^\infty dt\,|g_\pm(t)|^2 =\frac{1}{2\Gamma_\mathcal{B}} \left(\frac{1}{1-y_\mathcal{B}^2}\pm\frac{1}{1+x_\mathcal{B}^2}\right)\, , \qquad |y_\mathcal{B}|<1 \, .
 \label{eq:intro-integrals}
\end{equation}

The integrated wrong-sign to right-sign ratio is consequently
\begin{equation}
    r_\mathcal{B} =\frac{N_{\mathrm{WS}}}{N_{\mathrm{RS}}} =\left|\frac{q_\mathcal{B}}{p_\mathcal{B}}\right|^2 \frac{ x_\mathcal{B}^2+ y_\mathcal{B}^2}{2+x_\mathcal{B}^2-y_\mathcal{B}^2}\, .
 \label{eq:intro-ratio}
\end{equation}

In the CP-conserving limit, the fraction of wrong-sign events within this pair of tag channels is
\begin{equation}
    \chi_\mathcal{B}\equiv\frac{N_{\mathrm{WS}}}{N_{\mathrm{WS}}+N_{\mathrm{RS}}} =\frac{ x_\mathcal{B}^2+y_\mathcal{B}^2}{2(1+x_\mathcal{B}^2)}\, ,\qquad r_\mathcal{B}\simeq \chi_\mathcal{B} \simeq \frac{x_\mathcal{B}^2+ y_\mathcal{B}^2}{2} \quad (|x_\mathcal{B}| ,|y_\mathcal{B}| \ll1) \, .
 \label{eq:intro-mixing-rate}
\end{equation}

These ratios should be distinguished from the absolute probability of a particular wrong-sign decay, $\Gamma_{\mathrm{tag}} |q_\mathcal{B} /p_\mathcal{B}|^2 I_-$. Nonuniform acceptance requires inserting the measured efficiency into the time integrals. At small mixing, a wrong-sign search is sensitive to $x_\mathcal{B}^2+ y_\mathcal{B}^2$, with an additional factor $|q_\mathcal{B}/ p_\mathcal{B}|^2$ when CP violation in mixing is allowed, rather than to $\delta m_\mathcal{B}$ alone.

Standard Model weak interactions connect baryons with different flavour content, allowing heavy-baryon mass mixing to induce mixing in the light-baryon sector. Experimental limits on light-baryon oscillations can therefore provide indirect constraints on heavy-flavour baryon-number violation. These constraints probe the dispersive mass insertion and they do not, by themselves, determine the absorptive contribution to heavy-baryon mixing. In the following subsection, we introduce the light-sector inputs and develop the hadronic framework used to translate them into bounds on heavy-baryon mass mixing.

\subsection{Indirect constraints from \texorpdfstring{$|\Delta B|=2$}{Delta B=2} mixing}
\label{sec:deltab2}

We use neutron--antineutron and $\Lambda-\bar\Lambda$ mixing as probes of heavy-baryon mass insertions. We first summarise the light-sector constraints that serve as inputs to our analysis, before describing how Standard Model weak transitions relate them to heavy-baryon mixing.

The Super-Kamiokande search for intranuclear $n-\bar n$ transitions implies a free-neutron oscillation time $\tau_{n-\bar n}>4.7\times10^8\,{\rm s}$ at $90\%$ C.L.~\cite{Super-Kamiokande:2020bov}. Using the relation between the oscillation time and the mass insertion introduced above, this corresponds to $|\delta m_n|<1.4\times10^{-33}\,{\rm GeV}$. In the strange sector, BESIII has directly searched for $\Lambda-\bar\Lambda$ oscillations, obtaining $|\delta m_\Lambda|<2.1\times10^{-18}\,{\rm GeV}$ at $90\%$ C.L.~\cite{BESIII:2024gcd}. Substantially stronger indirect constraints can, however, be inferred from the Super-Kamiokande limit $\tau(pp\to K^+K^+)>1.7\times10^{32}\,{\rm yr}$ at $90\%$ C.L.~\cite{Super-Kamiokande:2014hie}, using hadronic estimates~\cite{Aitken:2017wie,Goity:1994dq,Beneito:2025ond,BasiBeneito:2026feb}. The analysis of Ref.~\cite{Aitken:2017wie} yielded $|\delta m_\Lambda|\lesssim10^{-30}\,{\rm GeV}$, whereas Refs.~\cite{Beneito:2025ond,BasiBeneito:2026feb} obtained the stronger estimate $|\delta m_\Lambda|\lesssim10^{-32}\,{\rm GeV}$, which we adopt. Our reference inputs are therefore
\begin{equation}
\begin{aligned}
    |\delta m_n| &< 1.4\times10^{-33}\,{\rm GeV}\,,\\
    |\delta m_\Lambda| &\lesssim 10^{-32}\,{\rm GeV}\,.
\end{aligned}
\label{eq:light_baryon_inputs}
\end{equation}
The adopted $\Lambda-\bar\Lambda$ bound is inferred from dinucleon decay and depends on the hadronic and nuclear approximations used in its extraction. The neutron bound is derived from an intranuclear oscillation search and also involves a nuclear conversion to the corresponding free-neutron oscillation time.

Suppose that Standard Model weak interactions induce a transition between a heavy neutral baryon $\mathcal{B}_H$ and a lighter neutral baryon $\mathcal{B}_L$, described at the hadronic level by
\begin{equation}
    {\cal L}_{\rm weak}^{\rm eff} = \bar {\mathcal B}_L \left(g^{\rm PC}_{HL}+ g^{\rm PV}_{HL}\gamma_5 \right) \mathcal{B}_H +\mathrm{h.c.} \, .
\end{equation}

\begin{figure*}[!htbp]
\centering
\begin{tabular}{cc}
\includegraphics[width=0.45\textwidth]{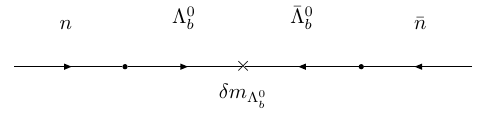}
&\hspace{+0.5cm}
\includegraphics[width=0.45\textwidth]{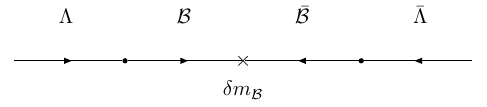}
\end{tabular}
\caption{Schematic hadron-level diagrams illustrating the indirect sensitivities considered in this work. The left panel shows the contribution of $\Lambda_b^0-\bar\Lambda_b^0$ mixing to $n-\bar n$ mixing, while the right panel shows the analogous contribution of $\mathcal B-\bar {\mathcal B}$ mixing to $\Lambda-\bar\Lambda$ mixing, with $\mathcal B=\Xi_b^0,\,\Xi_c^0,\,\Omega_c^0$. The crosses denote the corresponding $|\Delta B|=2$ baryon--antibaryon mass insertions, $\delta m_{\Lambda_b^0}$ and $\delta m_{\mathcal B}$, and the dots denote the effective weak-transition vertices associated with $g^{\rm PC}_{HL}$.}
\label{fig:transitions_hadronic}
\end{figure*}

For the parity-conserving contribution, inserting a $\mathcal{B}_H-\bar {\mathcal B}_H$ mass-mixing term between two such weak transitions induces, at the $\mathcal{B}_L$ pole,
\begin{equation}
    |\delta m_{\mathcal{B}_L}^{\rm ind}| \simeq \left| \frac{g^{\rm PC}_{HL}} {m_{\mathcal{B}_H}- m_{\mathcal{B}_L}} \right|^2 |\delta m_{\mathcal{B}_H}| \equiv R_{\mathcal{B}_H\mathcal{B}_L}\,|\delta m_{\mathcal{B}_H}| \, .
\label{eq:master}
\end{equation}
The parity-violating contribution instead involves the mass combination $m_{\mathcal{B}_H}+m_{\mathcal{B}_L}$ and therefore lacks the modest kinematic enhancement associated with the mass-difference denominator. In the CP-conserving limit, the two parity structures contribute without mixed PC--PV terms, giving 
\begin{equation}
    R_{\mathcal{B}_H\mathcal{B}_L}^{\rm PC+PV}=R_{\mathcal{B}_H\mathcal{B}_L}+ |g^{\rm PV}_{HL} /(m_{\mathcal{B}_H} +m_{\mathcal{B}_L})|^2 \, . 
\end{equation}
We retain only the parity-conserving contribution for our benchmark estimates. Within the hadronic approximation adopted here, and neglecting possible cancellations with other contributions to $\mathcal{B}_L-\bar {\mathcal B}_L$ mixing, an experimental upper limit on $|\delta m_{\mathcal{B}_L}|$ can be translated into the estimated indirect bound
\begin{equation}
    |\delta m_{\mathcal{B}_H}| \lesssim \frac{|\delta m_{\mathcal{B}_L}|}{R_{\mathcal{B}_H\mathcal{B}_L}} \,.
\label{eq:masterbound}
\end{equation}
This relation provides the organising principle of our analysis and is represented schematically at the hadronic level in Figure~\ref{fig:transitions_hadronic}. In particular, we take $\mathcal{B}_L=n$ for $\mathcal{B}_H= \Lambda_b^0$ and $\mathcal{B}_L =\Lambda$ for $\mathcal{B}_H= \Xi_b^0,\, \Xi_c^0,\,\Omega_c^0$.

\begin{figure*}[!htbp]
\centering
\begin{tabular}{cc}
\includegraphics[width=0.48\textwidth]{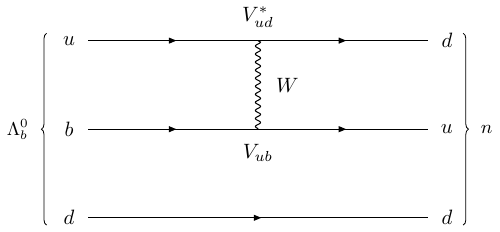}
&
\includegraphics[width=0.48\textwidth]{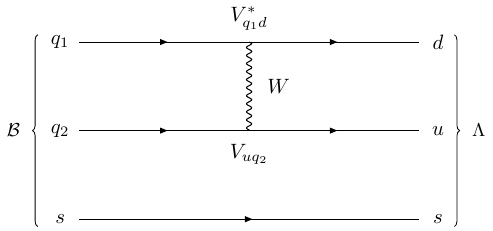}
\end{tabular}
\caption{Representative quark-level weak processes underlying the hadronic transitions of Figure~\ref{fig:transitions_hadronic}. The left panel corresponds schematically to the weak transition relevant for $\Lambda_b^0 \to n$, while the right panel illustrates the analogous transitions relevant for $\mathcal B\to\Lambda$ with $\mathcal B=\Xi_b^0,\Xi_c^0,\Omega_c^0$. For the right-hand diagram, $(q_1,q_2)=(u,b),(c,d),(c,s)$ correspond respectively to $\mathcal B=\Xi_b^0,\Xi_c^0,\Omega_c^0$, with the strange quark acting as a spectator. At low energies, these weak processes are encoded by local four-quark operators in the effective weak Hamiltonian.}
\label{fig:transitions_quark}
\end{figure*}

At the quark level, the weak transitions entering the hadronic diagrams are induced by flavour-changing Standard Model interactions, schematically illustrated in Figure~\ref{fig:transitions_quark}. Below the electroweak scale, these processes are described by local $|\Delta B|=0$ four-quark operators in the weak effective Hamiltonian, which mediate the required quark-flavour rearrangements. The quark-level diagrams are shown only as a guide to the underlying flavour structure.

Our approach isolates the contribution to light-baryon mixing mediated by hadronic weak transitions, treating the heavy-baryon mass-mixing parameter as an input without specifying the underlying $|\Delta B|=2$ operator. Additional short-distance electroweak contributions can arise at the quark level, depending on the flavour and chiral structure of the six-quark operators. These operator-dependent effects are not included in our analysis.

In principle, the weak baryon transitions illustrated at the hadronic level in Figure~\ref{fig:transitions_hadronic} could be computed directly from the hadronic matrix elements of the corresponding nonleptonic four-quark weak operators. However, these matrix elements are not currently available for the transitions considered here. We therefore adopt an intermediate-state approximation, in which selected contributions to the weak-transition amplitudes are estimated using semileptonic baryon form factors, as described in the following subsection.

\subsection{Nonleptonic heavy-to-light baryon transitions} \label{sec:nonleptonicframework}

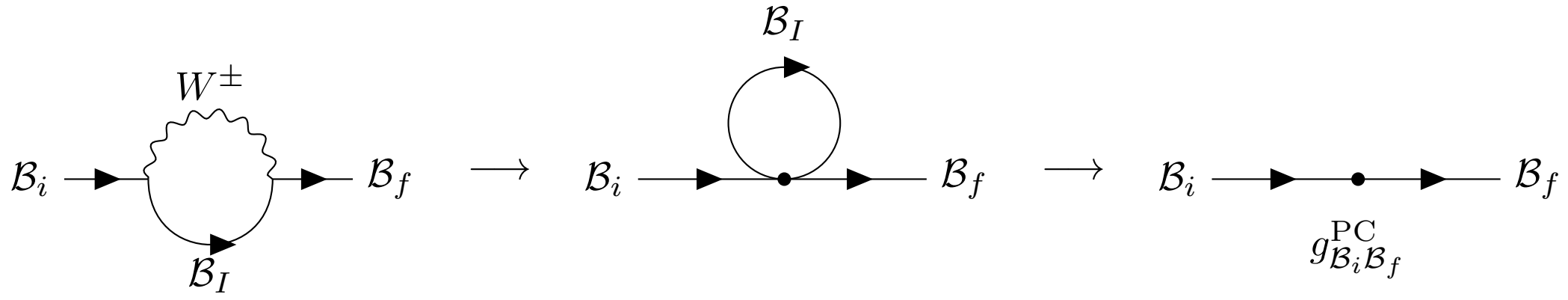
\begin{figure}[!htbp]
\centering
\begin{tabular}{@{}c@{\hspace{0.35cm}$\longrightarrow$\hspace{0.35cm}}c@{\hspace{0.35cm}$\longrightarrow$\hspace{0.35cm}}c@{}}

\begin{tikzpicture}[scale=1.1, baseline={(0,0)}]
\begin{feynman}
\vertex (i)  at (-1.45,0) {$\mathcal{B}_i$};
\vertex (v1) at (-0.50,0);
\vertex (v2) at (0.50,0);
\vertex (f)  at (1.45,0) {$\mathcal{B}_f$};

\diagram*{
 (i) -- [fermion] (v1),
 (v2) -- [fermion] (f),
 (v1) -- [boson, out=90, in=90, looseness=1.8,
           edge label=$W^\pm$] (v2),
 (v2) -- [anti fermion, out=-90, in=-90,
           looseness=1.8, edge label=$\mathcal{B}_I$] (v1),
};
\end{feynman}
\end{tikzpicture}

&

\begin{tikzpicture}[scale=1.1, baseline={(0,0)}]
\begin{feynman}
\vertex (i) at (-1.45,0) {$\mathcal{B}_i$};
\vertex[dot, minimum size=3pt, inner sep=0pt]
  (v) at (0,0) {};
\vertex (f) at (1.45,0) {$\mathcal{B}_f$};

\diagram*{
 (i) -- [fermion] (v),
 (v) -- [fermion] (f),
};

\draw[anti fermion] (v)
  arc[start angle=-90, end angle=270, radius=0.45cm];

\node at (0,1.25) {$\mathcal{B}_I$};

\end{feynman}
\end{tikzpicture}

&

\begin{tikzpicture}[scale=1.1, baseline={(0,0)}]
\begin{feynman}
\vertex (i) at (-1.45,0) {$\mathcal{B}_i$};
\vertex[dot, minimum size=3pt, inner sep=0pt]
  (v) at (0,0) {};
\vertex (f) at (1.45,0) {$\mathcal{B}_f$};

\diagram*{
 (i) -- [fermion] (v),
 (v) -- [fermion] (f),
};

\node[below=7pt] at (v) {$g^{\rm PC}_{\mathcal{B}_i \mathcal{B}_f}$};

\end{feynman}
\end{tikzpicture}

\end{tabular}

\caption{Schematic representation of the weak transition $\mathcal{B}_i\to \mathcal{B}_f$. The left diagram shows the baryonic self-energy mediated by an intermediate baryon $\mathcal{B}_I$ and a $W^\pm$ boson. In the low-energy limit, integrating out the $W$ yields a local Fermi interaction and a tadpole-like intermediate-baryon contraction (middle). The right diagram represents the corresponding effective parity-conserving weak coupling $g^{\rm PC}_{\mathcal{B}_i\mathcal{B}_f}$. The middle diagram is a schematic representation of the same contribution, rather than an additional one.}

\label{fig:weak_selfenergy}
\end{figure}

The weak coupling $g^{\rm PC}_{\mathcal{B}_i\mathcal{B}_f}$, with $\mathcal{B}_i= \mathcal{B}_H$ and $\mathcal{B}_f =\mathcal{B}_L$ (so that $g^{\rm PC}_{\mathcal{B}_i\mathcal{B}_f} \equiv g^{\rm PC}_{HL}$), is estimated by saturating the off-diagonal baryonic self-energy with selected low-lying intermediate baryons. At energies well below the $W$-boson mass, the corresponding diagram admits a tadpole-like representation in terms of a local Fermi interaction.
The parity-conserving component of the resulting transition amplitude is encoded in the effective weak vertex, as illustrated in Figure~\ref{fig:weak_selfenergy}.

Denoting by $p$ the external baryon momentum and by $k$ the momentum of the intermediate baryon $\mathcal{B}_I$, we define $q=p-k$ as the momentum transfer at the $\mathcal{B}_i\to \mathcal{B}_I$ vertex, with $-q=k-p$ at the $\mathcal{B}_I\to \mathcal{B}_f$ vertex. Retaining only the metric part of the $W$ propagator, the corresponding off-diagonal self-energy reads
\begin{equation}
    i\Sigma_{\mathcal{B}_f\mathcal{B}_i}(\slashed p)= -\frac{G_F M_W^2}{\sqrt{2}}\, \lambda_{\rm CKM} \int\frac{d^4k}{(2\pi)^4}\, \frac{ \Gamma^\mu_{ \mathcal{B}_I\to \mathcal{B}_f}(-q)(\slashed k+M_{\mathcal{B}_I}) \Gamma_{\mathcal{B}_i \to \mathcal{B}_I,\mu}(q)}{(k^2-M_{\mathcal{B}_I}^2+i\epsilon)(q^2-M_W^2+i\epsilon)}\,.
\label{eq:selfenergy}
\end{equation}
Here $\lambda_{\rm CKM}$ denotes the product of the CKM factors associated with the two charged-current vertices. Since only $|g^{\rm PC}_{ \mathcal{B}_i\mathcal{B}_f}|$ enters Eq.~\eqref{eq:master}, its overall phase convention is unphysical in our simplified approach. The weak vertices $\Gamma^\mu$ are constructed from the vector and axial-vector form factors of the corresponding semileptonic transitions,
\begin{align}
    \Gamma^\mu(q) ={}& f_1^V\gamma^\mu -f_2^V\frac{i\sigma^{\mu\nu}q_\nu}{M_{\rm ref}} +f_3^V\frac{q^\mu}{M_{\rm ref}} - \left[ f_1^A\gamma^\mu -f_2^A\frac{i\sigma^{\mu\nu}q_\nu}{M_{\rm ref}} +f_3^A\frac{q^\mu}{M_{\rm ref}}\right]\gamma_5 \, .
\label{eq:weakvertex}
\end{align}
Here $f_i^{V,A}\equiv f_i^{V,A}(q^2)$ and $\sigma^{\mu\nu}=\frac{i}{2}[\gamma^\mu, \gamma^\nu] $. We adopt the parent-baryon mass as the reference mass at each vertex, i.e. $M_{\rm ref}=M_{\mathcal{B}_i}$ at the $\mathcal{B}_i\to \mathcal{B}_I$ vertex and $M_{\rm ref}=M_{\mathcal{B}_I}$ at the $\mathcal{B}_I\to \mathcal{B}_f$ vertex. At each vertex, the momentum transfer is defined as the initial minus the final baryon momentum. All form factors are converted to the convention of Eq.~\eqref{eq:weakvertex}. The form-factor inputs and their conversion to this common convention are detailed in Appendix~\ref{app:formfactors}. In particular, the form factors for $\Lambda_b^0\to\Lambda_c^+$, $\Lambda_b^0\to p$, $\Lambda_c^+\to n$, $\Lambda_c^+\to\Lambda$, and $\Xi_c^0 \to\Xi^-$ are provided in the helicity basis and converted to the Dirac basis used here. Whenever available, we use the parametrisations adopted in the corresponding form-factor calculations. These are typically expressed through a pole-improved $z$ expansion, with the conformal variable $z(q^2)$ defined in Eq.~\eqref{eq:zdef},
\begin{equation}
    f_i^X(q^2)= \frac{1}{1-q^2/(M_{{\rm pole},i}^{X})^2} \sum_n a_{ni}^X z(q^2)^n\, , \qquad X=V,A\, .
\end{equation}

After performing the Dirac algebra, we isolate the parity-conserving scalar contribution to the off-diagonal self-energy. For our benchmark estimate, we neglect the external momentum in the loop kernel, $p\simeq 0$, so that $q=p-k\simeq-k$. The resulting coupling provides a zero-external-momentum approximation to the effective weak vertex entering Eq.~\eqref{eq:master}. For the numerical evaluation, we Wick-rotate the loop momentum, $k^2\to-k_E^2\equiv-K^2$. Since the semileptonic form factors are known only in their physical kinematic regions, their use inside the loop requires an extrapolation to off-shell and spacelike momenta. We adopt the Euclidean prescription
\begin{equation}
    q^2=(p-k)^2\longrightarrow -K^2,
\end{equation}
which neglects the external-momentum and angular dependence of the form factors. The remaining Euclidean angular integration is then
\begin{equation}
    \int\frac{d^4k_E}{(2\pi)^4}F(k_E^2)=\frac{1}{16\pi^2} \int_0^\infty dK^2\,K^2F(K^2) \, ,
\end{equation}
and the coupling can be written schematically as
\begin{equation}
    g^{\rm PC}_{\mathcal{B}_i\mathcal{B}_f}(\Lambda_{\rm cutoff}) = \frac{G_FM_W^2}{\sqrt{2}}\frac{\lambda_{\rm CKM}}{16\pi^2}\int_0^{\Lambda_{\rm cutoff}^2}dK^2\,\frac{K^2\,\mathcal N_{\rm PC}(K^2)}{\left( K^2+ M_{ \mathcal{B}_I }^2\right) \left(K^2+M_W^2\right)} \, .
\label{eq:genericloop}
\end{equation}
Here, $\Lambda_{\rm cutoff}$ is a UV cutoff on the Euclidean loop momentum. In principle, the cutoff is not required, as the integral in Eq.~(\ref{eq:genericloop}) is convergent as $K^2 \to \infty$. However, the convergence is naively rather slow. Because the process involves momentum transfer between the quarks inside the baryons, a perturbative gluon propagator should bring additional factors of $1/K^2$ at higher momentum transfers, improving the convergence. We feel that, rather than introducing somewhat ad hoc factors of $1/K^2$, a hadronic-scale cutoff would be more appropriate. As a benchmark, we take $\Lambda_{\rm cutoff}=10~{\rm GeV}$, which we consider a reasonable choice for the few-GeV baryon masses involved. The scalar numerator $\mathcal N_{\rm PC}$ is obtained by projecting the loop kernel onto the identity Dirac structure. In the $p\simeq 0$ approximation adopted above, it reads
\begin{equation}
    \mathcal N_{\rm PC}(K^2) = \left. \frac{1}{4}\operatorname{Tr} \left[ \Gamma^\mu_{\mathcal{B}_I\to \mathcal{B}_f}(k) (\slashed k+M_{\mathcal{B}_I}) \Gamma_{\mathcal{B}_i\to \mathcal{B}_I,\mu}(-k) \right] \right|_{k^2=-K^2}\, ,
\label{eq:NPC}
\end{equation}
where all form factors are evaluated at $q^2=-K^2$.

For an order-of-magnitude estimate, we retain only selected low-lying $J^P=1/2^+$ intermediate baryons for which the relevant hadronic transition form factors are available, neglecting excited states and other possible contributions. Our approach therefore isolates selected hadronic contributions rather than providing a complete calculation of the nonleptonic weak amplitudes. The cutoff dependence, the truncation of the intermediate-state spectrum, and the continuation of the form factors to off-shell and spacelike momenta constitute the dominant theoretical uncertainties. Our numerical results should accordingly be regarded as order-of-magnitude estimates rather than precision bounds. 

\section{Numerical results}
\label{sec:numerics}

We now apply Eqs.~\eqref{eq:master}--\eqref{eq:NPC}. The detailed form-factor input for every weak transition is collected in Appendix~\ref{app:formfactors}.

\subsection{\texorpdfstring{$\Xi_b^0-\bar\Xi_b^0$}{Xib osc}}

For $\Xi_b^0\to\Lambda$ we retain the ground-state $\Xi_c^+$ intermediate state, with $\lambda_{\rm CKM} = V_{cb}\,V_{cd}^*$. For the benchmark cutoff $\Lambda_{\rm cutoff}=10~{\rm GeV}$ we find
\begin{equation}
    |g^{\rm PC}_{\Xi_b^0\Lambda}| \simeq 5.3\times10^{-8}\, {\rm GeV}\, ,
\end{equation}
which gives
\begin{equation}
    R_{\Xi_b^0\Lambda} = \left| \frac{g^{\rm PC}_{\Xi_b^0\Lambda}}{m_{\Xi_b^0}-m_\Lambda}\right|^2\simeq 1.3\times10^{-16}\,.
\end{equation}

We also consider the contribution of the $\Sigma^+$ intermediate state, arising from the weak transitions $\Xi_b^0\to\Sigma^+$ and $\Sigma^+\to\Lambda$, with $\lambda_{\rm CKM}=V_{ub}V_{ud}^*$. Although these CKM factors enter the unitarity relation $\sum_{q=u,c,t} V_{qb}V_{qd}^*=0$, the two channels involve distinct intermediate baryons and hadronic form factors, so CKM unitarity does not enforce a GIM cancellation between them. The corresponding loop amplitude is evaluated using Eq.~\eqref{eq:genericloop} with $\mathcal{B}_I=\Sigma^+$. At the benchmark cutoff, its magnitude is approximately $10\%$ of the $\Xi_c^+$ amplitude, and we omit it from the quoted benchmark estimate. Using Eq.~\eqref{eq:light_baryon_inputs}, we obtain the estimated indirect bound
\begin{equation}
    |\delta m_{\Xi_b^0}| \lesssim 7.7\times10^{-17}\,{\rm GeV}\,,
    \label{eq:Xib-bound}
\end{equation}
almost three orders of magnitude below the direct LHCb limit quoted in Table~\ref{tab:heavy_baryon_osc}. An early proposal for rapid $\Xi_b^0-\bar\Xi_b^0$ oscillations was made in Ref.~\cite{Kuzmin:1996sy}, where transition times of order $10^{-13}\,{\rm s}$ were conjectured in scenarios with baryon-number-violating interactions involving all three quark generations, while neutron-antineutron transitions could remain strongly suppressed. Such rapid oscillations are already excluded by the direct LHCb search~\cite{LHCb:2017vth}. The authors of Ref.~\cite{Aitken:2017wie} derived an indirect bound from dinucleon decay, $|\delta m_{\Xi_b^0}|\lesssim10^{-10}\,{\rm GeV}$, using order-of-magnitude estimates of the hadronic matrix elements. They also obtained a stronger but highly model-dependent bound from collider searches, $|\delta m_{\Xi_b^0}|\lesssim10^{-17}\,{\rm GeV}$, within a model involving a colour-triplet scalar and Majorana fermions.

\subsection{\texorpdfstring{$\Lambda_b^0-\bar\Lambda_b^0$}{Lambdab osc}}

For $\Lambda_b^0\to n$ we retain the ground-state $\Lambda_c^+$ intermediate state, with $\lambda_{\rm CKM} = V_{cb}\,V_{cd}^* $. We have also evaluated the contribution from a proton intermediate state, $\Lambda_b^0 \to p\to n$, with $\lambda_{\rm CKM}= V_{ub} V_{ud}^*$, using $\Lambda_b^0\to p$ form factors from Ref.~\cite{Detmold:2015aaa} and nucleon form factors constrained by the conserved vector current (CVC) and partially conserved axial current (PCAC) relations, as detailed in Appendix~\ref{app:formfactors}. Despite the CKM ratio $|V_{ub}V_{ud}^* /(V_{cb} V_{cd}^*)|\simeq0.39$, including this contribution coherently changes $|g^{\rm PC}_{\Lambda_b^0n}|$ by less than $0.3\%$ at $\Lambda_{\rm cutoff}=10~{\rm GeV}$, and we therefore omit it from the quoted estimate. As in the $\Xi_b^0$ case, CKM unitarity does not enforce a GIM cancellation. For the benchmark cutoff $\Lambda_{\rm cutoff}=10~{\rm GeV}$ we obtain
\begin{equation}
    |g^{\rm PC}_{\Lambda_b^0n}| \simeq 1.5\times10^{-8}\,{\rm GeV}\,,
\end{equation}
which gives
\begin{equation}
R_{\Lambda_b^0n} = \left| \frac{g^{\rm PC}_{\Lambda_b^0n}}{m_{\Lambda_b^0}-m_n}\right|^2\simeq 1.0\times10^{-17}\,.
\end{equation}

Using Eq.~\eqref{eq:light_baryon_inputs}, we obtain the indirect sensitivity estimate
\begin{equation}
    |\delta m_{\Lambda_b^0}| \lesssim 1.4\times10^{-16}\, {\rm GeV}\,.
\label{eq:Lb-bound}
\end{equation}

We have also evaluated the $\Lambda_b^0\to\Lambda_c^+\to\Lambda$ transition, with $\lambda_{\rm CKM}=V_{cb}V_{cs}^*$, using the $\Lambda_c^+\to\Lambda$ lattice-QCD form factors of Ref.~\cite{Meinel:2016dqj}. At $\Lambda_{\rm cutoff}=10~{\rm GeV}$, this yields an indirect sensitivity $|\delta m_{\Lambda_b^0}|\lesssim1.7\times10^{-16}\,{\rm GeV}$, comparable to Eq.~\eqref{eq:Lb-bound}. We retain the neutron channel as our reference estimate because its light-baryon input follows from a dedicated search for intranuclear $n-\bar n$ transitions, whereas the $\Lambda$ input requires an additional hadronic estimate relating dinucleon decay to $\Lambda-\bar\Lambda$ mixing.

\subsection{\texorpdfstring{$\Xi_c^0-\bar\Xi_c^0$}{Xic osc}}
For $\Xi_c^0\to\Lambda$, we retain both the $\Xi^-$ and $\Sigma^-$ intermediate states, whose corresponding CKM factors are
\begin{equation}
\begin{aligned}
    \lambda^{\Xi^-}_{\rm CKM}
    &=V_{cs}^*V_{us}\simeq+0.21836,\\
    \lambda^{\Sigma^-}_{\rm CKM}
    &=V_{cd}^*V_{ud}\simeq-0.21857.
\end{aligned}
\end{equation}
Unlike in the previous cases, the two CKM factors have nearly equal magnitudes and opposite signs. We include both intermediate-state contributions coherently in the effective weak-transition amplitude. The $\Xi_c^0\to\Xi^-$ and $\Xi^-\to\Lambda$ vertices are described by lattice-QCD and QCD-sum-rule form factors, respectively, and the $\Xi_c^0\to\Sigma^-$ vertex by light-cone sum-rule form factors. For the $\Sigma^-\to\Lambda$ vertex, we adopt a simplified leading-order SU(3) approximation, retaining only the constant axial-vector form factor $g_1^{\Sigma\Lambda}=\sqrt{2/3}\,D$~\cite{Jenkins:1990jv}.

For the benchmark cutoff $\Lambda_{\rm cutoff}=10~{\rm GeV}$, the two contributions partially cancel, yielding
\begin{equation}
    |g_{\Xi_c^0\Lambda}^{\rm PC}|
    \simeq 1.1\times10^{-7}\,{\rm GeV}\,,
\end{equation}
which gives
\begin{equation}
    R_{\Xi_c^0\Lambda}
    =\left|
    \frac{g^{\rm PC}_{\Xi_c^0\Lambda}}
    {m_{\Xi_c^0}-m_\Lambda}
    \right|^2
    \simeq 6.6\times10^{-15}\,.
\end{equation}

Using the light-baryon input of Eq.~\eqref{eq:light_baryon_inputs}, we obtain the indirect sensitivity estimate
\begin{equation}
    |\delta m_{\Xi_c^0}|
    \lesssim 1.5\times10^{-18}\,{\rm GeV}\,.
    \label{eq:Xic-bound}
\end{equation}
The cancellation makes this estimate particularly sensitive to the relative sizes of the two hadronic contributions and their form-factor inputs.
For comparison, the direct Belle limit is
$|\delta m_{\Xi_c^0}|<5.1\times10^{-13}\,{\rm GeV}$~\cite{Belle:2024cab}.

\subsection{\texorpdfstring{$\Omega_c^0-\bar\Omega_c^0$}{Omegac osc}}

For $\Omega_c^0\to\Lambda$ we retain the $\Xi_c^+$ intermediate state, with $\lambda_{\rm CKM}=V_{us}V_{cd}^*$. An intermediate $\Xi^-$ state carries the same CKM factor and may yield a comparable contribution. However, its quantitative assessment requires additional assumptions about incompletely determined form factors and their momentum dependence, while the available $\Omega_c^0\to\Xi^-$ calculations exhibit substantial model dependence~\cite{Duan:2025aur}. We therefore use the $\Xi_c^+$ channel as a representative benchmark, and leave a combined treatment of both contributions beyond the scope of this work. For the benchmark cutoff $\Lambda_{\rm cutoff}=10~{\rm GeV}$ we obtain
\begin{equation}
    |g^{\rm PC}_{\Omega_c^0\Lambda}| \simeq 1.1\times10^{-8}\, {\rm GeV},
\end{equation}
which gives
\begin{equation}
    R_{\Omega_c^0\Lambda} = \left| \frac{g^{\rm PC}_{\Omega_c^0\Lambda}} {m_{\Omega_c^0}-m_\Lambda} \right|^2 \simeq 4.8\times10^{-17}\, .
\end{equation}

Using the indirect constraint in Eq.~\eqref{eq:light_baryon_inputs} on $\delta m_\Lambda$, we obtain the estimated indirect bound
\begin{equation}
    |\delta m_{\Omega_c^0}| \lesssim 2.1\times10^{-16}\, {\rm GeV} \,. \label{eq:Omegac-bound}
\end{equation}

\section{Summary and conclusions} \label{sec:summary}

In this work, we have investigated how Standard Model weak transitions can be exploited to derive indirect bounds on neutral baryon--antibaryon oscillations from existing limits on $n-\bar n$ oscillations and on $\Lambda-\bar\Lambda$ mixing. Our numerical results are summarised in Table~\ref{tab:generalbounds10Gev}, where we present the estimated indirect bounds on the mass-mixing parameters $|\delta m_\mathcal{B}|$ for the benchmark choice $\Lambda_{\rm cutoff}=10~{\rm GeV}$.

\begin{table}[!htbp]
\centering
\formfactortablestyle

\begin{tabular}{@{}lcc@{}}
\toprule
Oscillation
& $|\delta m_\mathcal{B}| \ {\rm [GeV]}$
& Light-baryon input \\
\midrule

$\Xi_b^0-\bar\Xi_b^0$
& $7.7\times10^{-17}$
& $|\delta m_\Lambda|$ \\

$\Lambda_b^0-\bar\Lambda_b^0$
& $1.4\times10^{-16}$
& $|\delta m_n|$ \\

$\Xi_c^0-\bar\Xi_c^0$
& $1.5\times10^{-18}$
& $|\delta m_\Lambda|$ \\

$\Omega_c^0-\bar\Omega_c^0$
& $2.1\times10^{-16}$
& $|\delta m_\Lambda|$ \\

$\Xi^0-\bar\Xi^0$
& $10^{-19}$
& $|\delta m_\Lambda|$ \\

\bottomrule
\end{tabular}

\caption{Estimated indirect bounds on the baryon--antibaryon mass-mixing parameters $|\delta m_\mathcal{B}|$, obtained from $|\delta m_n|<1.4\times10^{-33}\,{\rm GeV}$ and $|\delta m_\Lambda| \lesssim 10^{-32} \,{\rm GeV}$ (Eq.~\eqref{eq:light_baryon_inputs}). The heavy-baryon entries correspond to the benchmark cutoff $\Lambda_{\rm cutoff}=10~{\rm GeV}$ and are subject to the theoretical uncertainties of the hadronic weak-transition amplitudes discussed in Sec.~\ref{sec:framework}. The $\Xi^0$ entry is obtained in B$\chi$PT (Appendix~\ref{sec:Xiosc}) and does not depend on the cutoff.}
\label{tab:generalbounds10Gev}

\end{table}

Our main qualitative result is that Standard Model weak transitions can communicate baryon-number-violating interactions involving heavy or strange quarks to the light-baryon sector, allowing $n-\bar n$ oscillation and dinucleon-decay searches to provide indirect sensitivities substantially exceeding those of direct heavy-baryon searches. For $\Xi_b^0$ and $\Xi_c^0$, the estimated indirect bounds are roughly three and five orders of magnitude stronger than the direct LHCb and Belle limits, respectively.

Nevertheless, these results should be regarded as order-of-magnitude estimates rather than precise, model-independent bounds. An important source of uncertainty is the limited knowledge of the nonleptonic weak transitions, particularly for heavy baryons. Although nonleptonic light-hyperon decays have been studied extensively~\cite{Beneito:2025ond,Jenkins:1991bt,Borasoy:1998ku,Bijnens:1985kj}, the transitions considered here require phenomenological approximations involving hadronic amplitudes that do not generally correspond to physical on-shell decay processes.

A further limitation arises from the sizeable dependence of the loop integrals on the UV cutoff $\Lambda_{\rm cutoff}$. This sensitivity reflects the limited theoretical control over the continuation of semileptonic form factors into the off-shell and spacelike momentum regions probed by the loops. The cutoff should therefore be regarded as a phenomenological regulator rather than a physical parameter, and the benchmark results obtained for $\Lambda_{\rm cutoff}=10~{\rm GeV}$ should be interpreted accordingly. Partial cancellations between different contributions to the weak-transition amplitudes may further amplify these uncertainties.

The truncation of the intermediate-state spectrum introduces additional uncertainties. We retain only selected low-lying $J^P=1/2^+$ baryons, neglecting excited states, resonances, and multiparticle contributions. In particular, the bottom-baryon transitions involve intermediate charmed baryons, which may generate important long-distance effects, as also encountered in nonleptonic $B$-meson decays through the so-called ``charming penguins''~\cite{Ciuchini:1997hb,Ciuchini:1997rj,Ciuchini:2001gv}. The extent to which the selected intermediate states saturate the complete hadronic amplitudes remains unknown. Furthermore, our indirect constraints assume the absence of cancellations with independent $|\Delta B|=2$ contributions to light-baryon mixing.

These limitations motivate improved nonperturbative calculations of the relevant hadronic matrix elements. Lattice QCD has already achieved significant progress for the six-quark operators governing neutron--antineutron oscillations~\cite{Rinaldi:2018osy,Rinaldi:2019thf}, and extending these calculations to strange and heavy baryons would be particularly valuable. Dedicated lattice studies of nonleptonic weak transitions and long-distance hadronic contributions could also substantially reduce the uncertainties of our approach. Improved nuclear calculations would likewise help establish a more reliable connection between intranuclear baryon-number-violating processes and the underlying hadronic interactions.

Ultimately, our aim is to provide a phenomenological assessment of the potential sensitivity to heavy-baryon $|\Delta B|=2$ interactions. Despite the substantial theoretical uncertainties, our results illustrate how existing light-baryon constraints may probe baryon-number violation involving heavier quark flavours. Dedicated searches for strange- and heavy-baryon oscillations remain equally motivated, as different channels probe complementary flavour structures without relying on the same assumptions as the indirect approach. We hope that this study will encourage further experimental investigations and dedicated theoretical calculations, particularly from the lattice QCD community, towards a more comprehensive understanding of baryon-number violation across the baryon spectrum.

\acknowledgments

We thank V. O. Galkin for helpful correspondence concerning the form-factor parametrisations. Feynman diagrams were generated using the Ti\textit{k}Z-Feynman package for
\LaTeX~\cite{Ellis:2016jkw}. The work of ABB is funded by the grant CIACIF/2021/061 of the ``Generalitat Valenciana'' and also by the Spanish ``Agencia Estatal de Investigaci\'on'' through MICIN/AEI/10.13039/501100011033. SF acknowledges the financial support from the Slovenian Research Agency (research core funding No. P1-0035 and N1-0321). AAP was supported in part by the U.S. Department of Energy grant DE-SC0024357 and by the U.S. National Science Foundation grant PHY-2609993. 

\appendix

\section{Form-factor input}
\label{app:formfactors}

The numerical evaluation of the weak baryon transitions considered in this work requires hadronic form factors obtained from different theoretical approaches, including lattice QCD, relativistic quark models, and QCD sum rules. For completeness and reproducibility, we collect in this appendix the form-factor parametrisations and numerical inputs employed in our analysis. Table~\ref{tab:FFsummary} provides an overview of the relevant transitions, together with their corresponding references, theoretical methods, and form-factor parametrisations. The individual inputs are detailed in the following subsections, including the conversion between helicity and Dirac form factors whenever necessary.

We quote the independently determined fit parameters from the original references, accounting for the endpoint constraints where applicable. We also incorporate two minor corrections to the parametrisations of Refs.~\cite{Faustov:2018ahb,Faustov:2019ddj}.\footnote{We thank V.~O.~Galkin for confirming these typographical errors in private correspondence. In Ref.~\cite{Faustov:2018ahb}, the definition $t_+=(M_B+M_\pi)^2$ applies to $\Xi_b^- \to \Lambda$, whereas $t_+=(M_B+M_D)^2$ must be used for $\Xi_b^0\to\Xi_c^+$ to reproduce the reported form factors at $q^2=0$. Furthermore, in Table~II of Ref.~\cite{Faustov:2019ddj}, the coefficient $a_2$ of $f_1^A$ should read $-0.515$ instead of $-5.15$.}

\begin{table}[!htbp]
\centering
\formfactortablestyle

\begin{tabularx}{\linewidth}{@{}>{\raggedright\arraybackslash}p{2.0cm}>{\raggedright\arraybackslash}p{1.5cm}>{\raggedright\arraybackslash}p{3.0cm}>{\raggedright\arraybackslash}X@{}}
\toprule
Transition & Ref. & Method / basis & Form factors and parametrisation \\
\midrule

$\Xi_b^0\to\Xi_c^+$
&
\cite{Faustov:2018ahb}
&
relativistic quark model, Dirac basis
&
$f_{1,2,3}^{V,A}$, quadratic $z$ expansion with $a_0,a_1,a_2$
\\

$\Xi_b^0\to\Sigma^+$
&
\cite{Rui:2025bsu}
&
perturbative QCD, Dirac basis
&
$f_{1,2,3},g_{1,2,3}$, linear $z$ expansion with $a_0,a_1$
\\

$\Xi_c^+\to\Lambda$
&
\cite{Faustov:2019ddj}
&
relativistic quark model, Dirac basis
&
$f_{1,2,3}^{V,A}$, quadratic $z$ expansion with $a_0,a_1,a_2$
\\

$\Lambda_b^0\to\Lambda_c^+$
&
\cite{Detmold:2015aaa}
&
lattice QCD, helicity basis
&
$f_+,f_0,f_\perp,g_+,g_0,g_\perp$, linear $z$ expansion
\\

$\Lambda_b^0\to p$
&
\cite{Detmold:2015aaa}
&
lattice QCD, helicity basis
&
$f_+,f_0,f_\perp,g_+,g_0,g_\perp$, nominal linear $z$ expansion
\\

$p\to n$
&
\cite{LlewellynSmith:1971uhs,Galster:1971kv}
&
phenomenological (CVC, PCAC), Dirac basis
&
$f_{1,2}^V$ from dipole/Galster Sachs form factors, axial dipole $f_1^A$, pion-pole $f_3^A$
\\

$\Lambda_c^+\to n$
&
\cite{Meinel:2017ggx}
&
lattice QCD, helicity basis
&
$f_+,f_0,f_\perp,g_+,g_0,g_\perp$, quadratic $z$ expansion
\\

$\Lambda_c^+\to\Lambda$
&
\cite{Meinel:2016dqj}
&
lattice QCD, helicity basis
&
$f_+,f_0,f_\perp,g_+,g_0,g_\perp$, nominal quadratic $z$ expansion
\\

$\Xi_c^0\to\Xi^-$
&
\cite{Farrell:2025gis}
&
lattice QCD, helicity basis
&
$f_+,f_0,f_\perp,g_+,g_0,g_\perp$, cubic $z$ expansion with endpoint constraints
\\

$\Xi^-\to\Lambda$
&
\cite{Zhang:2024ick}
&
QCD sum rules, Dirac basis
&
$f_1(q^2),g_1(q^2)$ from $z$ fits and constant $f_2$
\\

$\Xi_c^0\to\Sigma^-$
&
\cite{Azizi:2011mw}
&
light-cone QCD sum rules, Dirac basis
&
$f_{1,2,3},g_{1,2,3}$, with double-pole fit
\\

$\Sigma^-\to\Lambda$
&
\cite{Jenkins:1990jv}
&
simplified leading-order SU(3) approximation
&
Constant $g_1=\sqrt{2/3}\,D$, remaining form factors neglected
\\

$\Omega_c^0\to\Xi_c^+$
&
\cite{Xing:2024okl}
&
light-front quark model, Dirac basis
&
$f_{1,2,3},g_{1,2,3}$, linear $z$ expansion with $a,b$
\\

\bottomrule
\end{tabularx}

\caption{Summary of the hadronic form-factor inputs employed in the calculation of the weak baryon transitions, indicating the theoretical approaches and momentum-transfer parametrisations adopted from the corresponding references.}
\label{tab:FFsummary}

\end{table}

\subsection{\texorpdfstring{$\Xi_b^0\to\Xi_c^+$}{XibXic}}\label{sec:appA1}

For this transition, we use the form factors of Ref.~\cite{Faustov:2018ahb}, defined through
\begin{align}
    \Gamma^\mu(q^2) ={}& f_1^V(q^2)\gamma^\mu -i f_2^V(q^2)\frac{\sigma^{\mu\nu}q_\nu}{M} +f_3^V(q^2)\frac{q^\mu}{M}  - \left[ f_1^A(q^2)\gamma^\mu -i f_2^A(q^2)\frac{\sigma^{\mu\nu}q_\nu}{M} +f_3^A(q^2)\frac{q^\mu}{M} \right]\gamma_5 \, ,
    \label{eq:A1parametrisation}
\end{align}
with $M=m_{\Xi_b^0}$. The published decay rates of Ref.~\cite{Faustov:2018ahb} are reproduced with $q=p_i-p_f$, so this decomposition coincides with Eq.~\eqref{eq:weakvertex} and the form factors of Refs.~\cite{Faustov:2018ahb,Faustov:2019ddj} enter without sign changes. They are given by the quadratic parametrisation
\begin{equation}
    F(q^2)=\frac{a_0+a_1z(q^2)+a_2z(q^2)^2} {1-q^2/M_{\rm pole}^2}\,. \label{eq:A2parametrisation}
\end{equation}
Here,
\begin{equation}
    z(q^2)=\frac{\sqrt{t_+-q^2}-\sqrt{t_+-t_0}}{\sqrt{t_+-q^2}+\sqrt{t_+-t_0}}\,, \qquad
    \begin{aligned}
        t_+&=(m_B+m_D)^2,\\
        t_0&=(m_{\Xi_b^0}-m_{\Xi_c^+})^2,
    \end{aligned}
\label{eq:zdef}
\end{equation}
where $m_B$ and $m_D$ denote the $B$- and $D$-meson masses, respectively. The numerical coefficients from Ref.~\cite{Faustov:2018ahb} are collected in Table~\ref{tab:XibXicFF}, and the pole masses are $6.333$ GeV for $f_{1,2}^V$,
$6.743$ GeV for $f_{1,2}^A$, $6.699$ GeV for $f_3^V$, and $6.275$ GeV for $f_3^A$.

\begin{table}[!htbp]
\centering
\formfactortablestyle

\begin{tabular}{@{}lrrrrrr@{}}
\toprule
&
$f_1^V$
&
$f_2^V$
&
$f_3^V$
&
$f_1^A$
&
$f_2^A$
&
$f_3^A$
\\
\midrule

$F(0)$
& 0.474 & 0.150 & 0.081 & 0.449 & $-0.030$ & $-0.285$ \\

$a_0$
& 0.684 & 0.308 & 0.121 & 0.729 & $-0.078$ & $-0.541$ \\

$a_1$
& $-5.16$ & $-4.18$ & $-0.315$ & $-7.11$ & 0.775 & 6.93 \\

$a_2$
& 28.0 & 25.9 & $-5.81$ & 41.5 & 0.372 & $-44.9$ \\

\bottomrule
\end{tabular}

\caption{Form-factor parameters for $\Xi_b^0\to\Xi_c^+$ from Ref.~\cite{Faustov:2018ahb}.}
\label{tab:XibXicFF}

\end{table}


\subsection{\texorpdfstring{$\Xi_b^0\to\Sigma^+$}{XibSigma}}
For the $\Xi_b^0\to\Sigma^+$ transition, Ref.~\cite{Rui:2025bsu} calculates the six vector and
axial-vector form factors $f_{1,2,3}$ and $g_{1,2,3}$ within perturbative QCD. Their $q^2$ dependence is parametrised using the linear $z$ expansion,
\begin{equation}
    F(q^2)=
    \frac{a_0+a_1 z(q^2)}{1-q^2/M_{\rm pole}^2}\,,
\end{equation}
where
\begin{equation}
    z(q^2) =
    \frac{\sqrt{Q_+^2-q^2}-\sqrt{Q_+^2-Q_-^2}}
    {\sqrt{Q_+^2-q^2}+\sqrt{Q_+^2-Q_-^2}}\,,
    \qquad
    Q_\pm=M_{\Xi_b^0}\pm M_{\Sigma^+}\,.
\end{equation}

We adopt the Exponential model (S1) for the $\Xi_b^0$ light-cone distribution amplitudes, with the corresponding form-factor parameters collected in Table~\ref{tab:XibSigmaFF}. Alternative parametrisations can be found in Ref.~\cite{Rui:2025bsu}.

\begin{table}[!htbp]
\centering
\formfactortablestyle
\begin{tabular}{@{}lrrrrrr@{}}
\toprule
Parameter
& $f_1$ & $f_2$ & $f_3$
& $g_1$ & $g_2$ & $g_3$ \\
\midrule
$F(0)$
& 0.200 & 0.052 & $-0.038$
& 0.213 & 0.046 & $-0.043$ \\
$a_0$
& 0.361 & 0.138 & $-0.109$
& 0.407 & 0.129 & $-0.117$ \\
$a_1$
& $-1.121$ & $-0.612$ & 0.505
& $-1.386$ & $-0.590$ & 0.519 \\
\bottomrule
\end{tabular}
\caption{Form-factor parameters for $\Xi_b^0\to\Sigma^+$ in the Exponential model (S1) of Ref.~\cite{Rui:2025bsu}.}
\label{tab:XibSigmaFF}
\end{table}

The pole masses used in the $z$ expansion are
\begin{equation}
    M_{\rm pole} =
    \begin{cases}
        5.325~{\rm GeV}, & f_{1,2},\\
        5.723~{\rm GeV}, & g_{1,2},\\
        5.749~{\rm GeV}, & f_3,\\
        5.280~{\rm GeV}, & g_3.
    \end{cases}
\end{equation}
For the subsequent $\Sigma^+ \to \Lambda$ transition, we use the isospin counterpart of the $\Sigma^-\to\Lambda$ vertex described in Appendix~\ref{app:SigmaLambda}, retaining only the constant axial-vector form factor $|g_1^{\Sigma^+\Lambda}|=\sqrt{2/3}\,D$.

\subsection{\texorpdfstring{$\Xi_c^+\to\Lambda$}{XicLambda}}

For the $\Xi_c^+\to\Lambda$ transition, we adopt the form factors of Ref.~\cite{Faustov:2019ddj}, defined as in Eq.~\eqref{eq:A1parametrisation} with $M=m_{\Xi_c^+}$ and, as discussed in Appendix~\ref{sec:appA1}, entering without sign changes. They are parametrised as in Eqs.~\eqref{eq:A2parametrisation}--\eqref{eq:zdef} with $t_+=(m_D+m_\pi)^2$ and $t_0=(m_{\Xi_c^+}-m_\Lambda)^2$, and the numerical parameters are collected in Table~\ref{tab:XicLambdaFF}. The corresponding pole masses are $2.010$ GeV for $f_{1,2}^V$, $2.423$ GeV for $f_{1,2}^A$ and $f_3^V$, and $1.870$ GeV for $f_3^A$.

\begin{table}[!htbp]
\centering
\formfactortablestyle

\begin{tabular}{@{}lrrrrrr@{}}
\toprule
&
$f_1^V$
&
$f_2^V$
&
$f_3^V$
&
$f_1^A$
&
$f_2^A$
&
$f_3^A$
\\
\midrule

$F(0)$
& 0.203 & 0.165 & 0.120 & 0.201 & $-0.054$ & $-0.427$ \\

$a_0$
& 0.166 & 0.202 & 0.179 & 0.189 & $-0.148$ & $-0.453$ \\

$a_1$
& 0.402 & $-0.142$ & $-0.091$ & 0.154 & 0.481 & $-0.084$ \\

$a_2$
& $-1.04$ & $-0.653$ & $-2.00$ & $-0.515$ & 0.917 & 1.69 \\

\bottomrule
\end{tabular}

\caption{Form-factor parameters for $\Xi_c^+\to\Lambda$ from Ref.~\cite{Faustov:2019ddj}.}
\label{tab:XicLambdaFF}

\end{table}


\subsection{\texorpdfstring{$\Lambda_b^0\to\Lambda_c^+$}{LambdabLambdac}}

For the $\Lambda_b^0\to\Lambda_c^+$ transition, we adopt the lattice-QCD form factors of Ref.~\cite{Detmold:2015aaa}, whose numerical parameters are collected in Table~\ref{tab:LbLcFF}. The conformal variable is defined with $t_0=(m_{\Lambda_b^0}-m_{\Lambda_c^+})^2$ and the form-factor-dependent choice $t_+^F=(M_{\rm pole}^F)^2$, following the conventions of Ref.~\cite{Detmold:2015aaa}. These form factors are given in the helicity basis and must be converted to the Dirac basis employed in our loop calculation.

For a generic transition $\mathcal B_i\to\mathcal B_f$,
with baryon masses $M=m_{\mathcal B_i}$ and $m=m_{\mathcal B_f}$ and momentum transfer $q=p_i-p_f$, as in Eq.~\eqref{eq:weakvertex}, the conversion between the helicity and Dirac bases reads
\begin{align}
f_1^V
&=
\frac{(M+m)^2f_+-q^2f_\perp}{s_+}, \label{eq:hel2dirac}
&
f_2^V
&=
M\frac{M+m}{s_+}(f_\perp-f_+),
\\
f_3^V
&=
\frac{M(M-m)}{q^2}(f_0-f_1^V),
&
f_1^A
&=
\frac{(M-m)^2g_+-q^2g_\perp}{s_-},
\\
f_2^A
&=
M\frac{M-m}{s_-}(g_+-g_\perp),
&
f_3^A
&=
\frac{M(M+m)}{q^2}(f_1^A-g_0),
\end{align}
where
\begin{equation}
    s_\pm=(M\pm m)^2-q^2\,.
\label{eq:spm}
\end{equation}

\begin{table}[!htbp]
\centering
\formfactortablestyle
\begin{tabular}{@{}lccc@{}}
\toprule
$F$
& $M_{\rm pole}$ [GeV]
& $a_0$
& $a_1$ \\
\midrule

$f_+$
& 6.332 & $0.8146\pm0.0167$ & $-4.8990\pm0.5425$ \\

$f_0$
& 6.725 & $0.7439\pm0.0125$ & $-4.6480\pm0.6084$ \\

$f_\perp$
& 6.332 & $1.0780\pm0.0256$ & $-6.4170\pm0.8480$ \\

$g_+$
& 6.768 & $0.6847\pm0.0086$ & $-4.4310\pm0.3572$ \\

$g_\perp$
& 6.768 & $0.6847\pm0.0086$ & $-4.4630\pm0.3613$ \\

$g_0$
& 6.276 & $0.7396\pm0.0143$ & $-4.3660\pm0.3314$ \\

\bottomrule
\end{tabular}
\caption{Lattice-QCD form-factor parameters for $\Lambda_b^0\to\Lambda_c^+$ from Ref.~\cite{Detmold:2015aaa}. The common $a_0$ for $g_+$ and $g_\perp$ implements the zero-recoil constraint.}
\label{tab:LbLcFF}
\end{table}


\subsection{\texorpdfstring{$\Lambda_b^0\to p$}{Lambdabp}}

For the $\Lambda_b^0\to p$ transition, we adopt the lattice-QCD helicity form factors of Ref.~\cite{Detmold:2015aaa}, using the nominal linear $z$ expansion
\begin{equation}
    F(q^2)= \frac{a_0^F+a_1^F z(q^2)}{1-q^2/(M_{\rm pole}^F)^2}\,, \qquad F=f_+,f_0,f_\perp,g_+,g_0,g_\perp\,.
\end{equation}

The conformal variable of Eq.~\eqref{eq:zdef} is defined with $t_0=(m_{\Lambda_b^0}-m_p)^2$ and $t_+=(m_B+ m_\pi )^2$ for all six form factors, using $m_B =5.279~{\rm GeV}$ and $m_\pi=0.135~{\rm GeV}$. The numerical parameters are collected in Table~\ref{tab:LbpFF}. The conversion to the Dirac basis follows Eqs.~\eqref{eq:hel2dirac}--\eqref{eq:spm}, with $M= m_{ \Lambda_b^0}$ and $m=m_p$.

\begin{table}[!htbp]
\centering
\formfactortablestyle
\begin{tabular}{@{}lccc@{}}
\toprule
$F$
& $M_{\rm pole}$ [GeV]
& $a_0$
& $a_1$ \\
\midrule

$f_+$
& 5.325 & $0.4382\pm0.0315$ & $-0.6452\pm0.2093$ \\

$f_0$
& 5.656 & $0.4189\pm0.0256$ & $-0.7862\pm0.2038$ \\

$f_\perp$
& 5.325 & $0.5389\pm0.0435$ & $-0.8069\pm0.3039$ \\

$g_+$
& 5.706 & $0.3912\pm0.0198$ & $-0.8167\pm0.1749$ \\

$g_\perp$
& 5.706 & $0.3912\pm0.0198$ & $-0.9061\pm0.1956$ \\

$g_0$
& 5.279 & $0.4526\pm0.0292$ & $-0.7817\pm0.1886$ \\

\bottomrule
\end{tabular}
\caption{Nominal lattice-QCD form-factor parameters for $\Lambda_b^0 \to p$ from Ref.~\cite{Detmold:2015aaa}. The common $a_0$ for $g_+$ and $g_\perp$ implements the zero-recoil constraint. The quoted uncertainties are those of the nominal fit.}
\label{tab:LbpFF}
\end{table}

\subsection{\texorpdfstring{$p\to n$}{Proton to neutron}}

For the $p\to n$ transition, we use a phenomenological dipole/Galster parametrisation of the vector form factors, an axial dipole, and pion-pole dominance for the induced pseudoscalar contribution~\cite{LlewellynSmith:1971uhs,Galster:1971kv}. Defining $Q^2=-q^2$, $\overline m_N=(m_p+m_n)/2$, $\tau= Q^2/(4\overline m_N^2)$, and $G_D= (1+Q^2/M_V^2)^{-2}$, the isovector Sachs form factors are
\begin{equation}
    G_E^{(v)}=G_D\left(1+\frac{\mu_n\tau}{1+5.6\tau} \right)\, , \qquad G_M^{(v)}=(\mu_p-\mu_n)G_D\, .
\end{equation}

In the convention of Eq.~\eqref{eq:weakvertex}, with $M_{\rm ref}=m_p$, the Dirac form factors evaluated at $q^2=-Q^2$ are
\begin{align}
    f_1^V&=\frac{G_E^{(v)}+\tau G_M^{(v)}}{1+\tau}\, ,
    &
    f_2^V&=\frac{m_p}{m_p+m_n}
           \frac{G_M^{(v)}-G_E^{(v)}}{1+\tau}\, ,
    &
    f_3^V&=0\, ,
    \\
    f_1^A&=\frac{g_A}{(1+Q^2/M_A^2)^2}\, ,
    &
    f_3^A&=-\frac{2m_p\overline m_N}
                     {m_{\pi^\pm}^2+Q^2}\,f_1^A\, ,
    &
    f_2^A&=0\, .
\end{align}
We take $M_V=0.84~{\rm GeV}$, $M_A=1.03~{\rm GeV}$, $g_A =1.2754$, $m_{\pi^\pm}=0.13957039~{\rm GeV}$, $\mu_p= 2.7928473446$, and $\mu_n=-1.91304273$, with magnetic moments expressed in nuclear magnetons.

\subsection{\texorpdfstring{$\Lambda_c^+\to n$}{Lambdacn}}

For the $\Lambda_c^+\to n$ transition, we use the lattice-QCD helicity form factors of Ref.~\cite{Meinel:2017ggx}, with the nominal quadratic $z$-expansion parameters collected in Table~\ref{tab:LcnFF}. The conformal variable is defined with $t_+=(m_D+m_\pi)^2$ and $t_0=(m_{\Lambda_c^+}-m_n)^2$, following Ref.~\cite{Meinel:2017ggx}. Their conversion to the Dirac basis follows Eqs.~\eqref{eq:hel2dirac}--\eqref{eq:spm}, with $M=m_{\Lambda_c^+}$ and $m=m_n$. The form factors satisfy the endpoint constraints
\begin{equation}
f_0(0)=f_+(0),
\qquad
g_0(0)=g_+(0),
\qquad
g_\perp(q^2_{\rm max})=g_+(q^2_{\rm max}).
\label{eq:endpoint}
\end{equation}

\begin{table}[!htbp]
\centering
\formfactortablestyle
\begin{tabular}{@{}lcccc@{}}
\toprule
$F$
& $M_{\rm pole}$ [GeV]
& $a_0$
& $a_1$
& $a_2$ \\
\midrule

$f_\perp$
& 2.010 & $1.36\pm0.07$ & $-1.70\pm0.83$ & $0.71\pm4.34$ \\

$f_+$
& 2.010 & $0.83\pm0.04$ & $-2.33\pm0.56$ & $8.41\pm3.05$ \\

$f_0$
& 2.351 & $0.84\pm0.04$ & $-2.57\pm0.60$ & $9.87\pm3.15$ \\

$g_\perp$
& 2.423 & $0.69\pm0.02$ & $-0.68\pm0.32$ & $0.70\pm2.18$ \\

$g_+$
& 2.423 & $0.69\pm0.02$ & $-0.90\pm0.29$ & $2.25\pm1.90$ \\

$g_0$
& 1.870 & $0.73\pm0.04$ & $-0.97\pm0.52$ & $0.83\pm2.61$ \\

\bottomrule
\end{tabular}
\caption{Nominal lattice-QCD form-factor parameters for $\Lambda_c^+\to n$ from Ref.~\cite{Meinel:2017ggx}.}
\label{tab:LcnFF}
\end{table}


\subsection{\texorpdfstring{$\Lambda_c^+\to\Lambda$}{LambdacLambda}}

For the $\Lambda_c^+\to\Lambda$ transition, we adopt the lattice-QCD helicity form factors of Ref.~\cite{Meinel:2016dqj}, using the nominal quadratic $z$ expansion
\begin{equation}
    F(q^2)= \frac{a_0^F+a_1^F z(q^2)+a_2^F[z(q^2)]^2}{1-q^2/(M_{\rm pole}^F)^2}\,, \qquad F=f_+,f_0,f_\perp,g_+,g_0,g_\perp\,.
\end{equation}

The conformal variable of Eq.~\eqref{eq:zdef} is defined with $t_0=(m_{\Lambda_c^+}-m_\Lambda)^2$ and $t_+=(m_D+m_K)^2$, using $m_D=1.870~{\rm GeV}$ and $m_K=0.494~{\rm GeV}$. The numerical parameters are collected in Table~\ref{tab:LcLambdaFF}. The conversion to the Dirac basis follows Eqs.~\eqref{eq:hel2dirac}--\eqref{eq:spm}, with $M=m_{\Lambda_c^+}$ and $m=m_\Lambda$.

\begin{table}[!htbp]
\centering
\formfactortablestyle
\begin{tabular}{@{}lcccc@{}}
\toprule
$F$
& $M_{\rm pole}$ [GeV]
& $a_0$
& $a_1$
& $a_2$ \\
\midrule

$f_+$
& 2.112 & $0.81\pm0.03$ & $-2.89\pm0.52$ & $7.82\pm4.53$ \\

$f_0$
& 2.318 & $0.77\pm0.02$ & $-2.24\pm0.51$ & $5.38\pm4.80$ \\

$f_\perp$
& 2.112 & $1.30\pm0.06$ & $-3.27\pm1.18$ & $7.16\pm11.6$ \\

$g_+$
& 2.460 & $0.68\pm0.02$ & $-2.44\pm0.25$ & $13.7\pm2.15$ \\

$g_\perp$
& 2.460 & $0.68\pm0.02$ & $-1.91\pm0.35$ & $6.24\pm4.89$ \\

$g_0$
& 1.968 & $0.71\pm0.03$ & $-2.86\pm0.44$ & $11.8\pm2.47$ \\

\bottomrule
\end{tabular}
\caption{Nominal lattice-QCD form-factor parameters for $\Lambda_c^+ \to\Lambda$ from Ref.~\cite{Meinel:2016dqj}. The common $a_0$ for $g_+$ and $g_\perp$ implements the zero-recoil constraint.}
\label{tab:LcLambdaFF}
\end{table}

\subsection{\texorpdfstring{$\Xi_c^0\to\Xi^-$}{Xic to Xi}}

For the $\Xi_c^0\to\Xi^-$ transition, we adopt the lattice-QCD helicity form factors of Ref.~\cite{Farrell:2025gis}, whose numerical parameters are collected in
Table~\ref{tab:XicXiFF}. In the physical limit, the form factors are parametrised as
\begin{equation}
    F(q^2) = \frac{a_0^F+a_1^Fz_F(q^2)+a_2^Fz_F(q^2)^2+a_3^Fz_F(q^2)^3}{1-q^2/(M_{\rm pole}^F)^2} \, .
\end{equation}

The conformal variable $z_F$ is defined as in Eq.~\eqref{eq:zdef}, with $t_0=(m_{\Xi_c^0}-m_{\Xi^-})^2$ and the form-factor-dependent
branch points
\begin{align}
t_+^{f_+,f_\perp,f_0}&=(m_D+m_K)^2,\\
t_+^{g_+,g_\perp,g_0}&=(m_{D^*}+m_K)^2.
\end{align}
Only independently determined coefficients are given numerically in Table~\ref{tab:XicXiFF}. Coefficients eliminated by endpoint constraints are marked ``constrained''.

\begin{table}[!htbp]
\centering
\formfactortablestyle

\begin{tabular}{@{}lcccccc@{}}
\toprule
$F$
&
$J^P$
&
$M_{\rm pole}$ [GeV]
&
$a_0$
&
$a_1$
&
$a_2$
&
$a_3$
\\
\midrule

$f_+$
& $1^-$
& 2.112
& $0.957\pm0.020$
& $-4.483\pm0.453$
& $16.22\pm5.17$
& $-20.33\pm28.79$
\\

$f_\perp$
& $1^-$
& 2.112
& $1.757\pm0.042$
& $-6.555\pm0.614$
& $12.56\pm6.07$
& $-6.574\pm29.63$
\\

$f_0$
& $0^+$
& 2.318
& $0.903\pm0.020$
& $-3.473\pm0.428$
& \textit{constrained}
& $-4.710\pm28.72$
\\

$g_+$
& $1^+$
& 2.460
& $0.764\pm0.014$
& $-3.455\pm0.271$
& $12.73\pm4.22$
& $7.531\pm28.40$
\\

$g_\perp$
& $1^+$
& 2.460
& $0.764\pm0.014$
& $-3.295\pm0.336$
& $11.46\pm4.90$
& $13.89\pm29.26$
\\

$g_0$
& $0^-$
& 1.968
& $0.836\pm0.017$
& $-5.143\pm0.380$
& \textit{constrained}
& $0.090\pm28.91$
\\

\bottomrule
\end{tabular}

\caption{Lattice-QCD form-factor parameters for $\Xi_c^0\to\Xi^-$ from Ref.~\cite{Farrell:2025gis}.}
\label{tab:XicXiFF}

\end{table}

The endpoint constraints are those of Eq.~\eqref{eq:endpoint}, with $q^2_{\rm max}=(m_{\Xi_c^0}-m_{\Xi^-})^2$. Since $z(q^2_{\rm max})=0$, the last condition implies $a_0^{g_+}=a_0^{g_\perp}$. The helicity-to-Dirac conversion of Eqs.~\eqref{eq:hel2dirac}--\eqref{eq:spm} is used with $M=m_{\Xi_c^0}$ and $m=m_{\Xi^-}$.

\subsection{\texorpdfstring{$\Xi^-\to\Lambda$}{Xi to Lambda}}

For the $\Xi^-\to\Lambda$ transition, we adopt the form factors of Ref.~\cite{Zhang:2024ick}, whose numerical parameters are collected in Table~\ref{tab:XiLambdaFF}. The $q^2$ dependence of $f_1$ and $g_1$ is parametrised as
\begin{equation}
    F(q^2)=\frac{F(0)}{1-q^2/M_{\rm pole}^2} \left[1+a_1\bigl(z(q^2,t_0)-z(0,t_0)\bigr)\right]\, ,
\end{equation}
where $z(q^2,t_0)$ is defined as in Eq.~\eqref{eq:zdef}, with
\begin{align}
    t_\pm&=(m_{\Xi^-}\pm m_\Lambda)^2,\qquad t_{\min}=-0.05~{\rm GeV}^2,\nonumber\\ t_0&=t_+-\sqrt{t_+-t_-}\sqrt{t_+-t_{\min}} \, .
\end{align}
The form factor $f_2$ is treated as constant. No independent input for $f_3$, $g_2$, or $g_3$ is used in our numerical implementation. The form factors in Table~\ref{tab:XiLambdaFF} are quoted in the convention of the original reference, which uses $+i\sigma^{\mu\nu}q_\nu f_2/m_{\Xi^-}$ with
$q=p_{\Xi^-}-p_\Lambda$. Matching to our convention in Eq.~\eqref{eq:weakvertex} therefore gives
\begin{equation}
    f_1^V=f_1,\qquad f_2^V=-f_2,\qquad f_1^A=g_1.
\end{equation}

\begin{table}[!htbp]
\centering
\formfactortablestyle

\begin{tabular}{@{}lccc@{}}
\toprule
$F$
&
$F(0)$
&
$a_1$
&
$M_{\rm pole}$ [GeV]
\\
\midrule

$f_1$
& $1.202\pm0.072$
& $-76.14\pm11.51$
& 0.892
\\

$g_1$
& $0.309\pm0.034$
& $-169.04\pm2.19$
& 1.270
\\

$f_2$
& $0.118\pm0.029$
& --
& --
\\

\bottomrule
\end{tabular}

\caption{Form-factor input for $\Xi^-\to\Lambda$ from Ref.~\cite{Zhang:2024ick}. The form factor $f_2$ is taken constant. The entries follow the convention of the original reference. In the convention of Eq.~\eqref{eq:weakvertex}, the tensor form factor enters as $f_2^V=-f_2$.}
\label{tab:XiLambdaFF}

\end{table}

\subsection{\texorpdfstring{$\Xi_c^0\to\Sigma^-$}{Xic to Sigma}}

For the $\Xi_c^0\to\Sigma^-$ transition, we adopt the light-cone QCD sum-rule form factors of Ref.~\cite{Azizi:2011mw}, whose numerical parameters are collected in Table~\ref{tab:XicSigmaFF}. In Ref.~\cite{Azizi:2011mw}, the tensor and induced terms are written as $+i\sigma^{\mu\nu}q_\nu f_2+q^\mu f_3$ (and analogously for $g_{2,3}$), without a mass factor and with $q=p_i-p_f$, so that $f_{2,3}$ and $g_{2,3}$ carry dimension GeV$^{-1}$. In the convention of Eq.~\eqref{eq:weakvertex} this gives $f_1^V=f_1$, $f_2^V=-m_{\Xi_c^0}f_2$, $f_3^V=m_{\Xi_c^0}f_3$, $f_1^A=g_1$, $f_2^A=-m_{\Xi_c^0}g_2$ and $f_3^A=m_{\Xi_c^0}g_3$. Their $q^2$ dependence is parametrised as
\begin{equation}
    F(q^2) = \frac{a}{1-q^2/M_{\rm pole}^2} + \frac{b}{(1-q^2/M_{\rm pole}^2)^2}\,.
\end{equation}

\begin{table}[!htbp]
\centering
\formfactortablestyle

\begin{tabular}{@{}lcccc@{}}
\toprule
$F$
& $a$
& $b$
& $M_{\rm pole}$ [GeV]
& $F(0)$ \\
\midrule

$f_1$
& 0.528
& $-0.119$
& 1.52
& $0.409\pm0.106$ \\

$f_2$
& $-0.564$
& 0.006
& 1.58
& $-0.558\pm0.145$ \\

$f_3$
& $-0.598$
& $-0.133$
& 1.57
& $-0.731\pm0.190$ \\

$g_1$
& 0.448
& $-0.031$
& 1.55
& $0.417\pm0.104$ \\

$g_2$
& $-0.493$
& 0.096
& 1.61
& $-0.397\pm0.099$ \\

$g_3$
& $-0.295$
& $-0.329$
& 1.63
& $-0.624\pm0.150$ \\

\bottomrule
\end{tabular}

\caption{Form-factor parameters for $\Xi_c\to\Sigma$ from Ref.~\cite{Azizi:2011mw} (Table~14 there), in the convention of that reference. $f_{2,3}$ and $g_{2,3}$ are in GeV$^{-1}$.}
\label{tab:XicSigmaFF}

\end{table}

\subsection{\texorpdfstring{$\Sigma^-\to\Lambda$}{Sigma to Lambda}} \label{app:SigmaLambda}

For the $\Sigma^-\to\Lambda$ transition, we adopt a simplified leading-order SU(3) approximation, retaining only the axial-vector form factor
\begin{equation}
    g_1^{\Sigma\Lambda}(q^2) =\sqrt{\frac{2}{3}}\,D\simeq0.656\,,
\end{equation}
with $D=0.804$~\cite{Jenkins:1990jv}. The remaining form factors are set to zero, and the momentum dependence of $g_1$ is neglected.

\subsection{\texorpdfstring{$\Omega_c^0\to\Xi_c^+$}{Omegac to Xic}}

For $\Omega_c^0\to\Xi_c^+$ we use the light-front quark-model form factors of Ref.~\cite{Xing:2024okl}, parametrised as
\begin{equation}
    F(q^2) = \frac{a+b\,z(q^2)}{1-q^2/M_{\rm pole}^2}\, ,
\end{equation}
with the corresponding parameters collected in Table~\ref{tab:OmegacXicFF}.

Here, $z(q^2)$ follows the definition given in Eq.~\eqref{eq:zdef}, with $t_+=(m_K+m_\pi)^2$ and $t_0=(m_{\Omega_c^0}-m_{\Xi_c^+})^2$, as prescribed in Ref.~\cite{Xing:2024okl}. In that reference, the tensor term is written as $+i\sigma^{\mu\nu}q_\nu f_2/M$ with $q=p_i-p_f$. Hence, in the convention of Eq.~\eqref{eq:weakvertex}, $f_{1,3}^V= f_{1,3}$, $f_2^V=-f_2$, $f_{1,3}^A= g_{1,3}$, and $f_2^A=-g_2$. The subsequent $\Xi_c^+\to\Lambda$ vertex is given in Table~\ref{tab:XicLambdaFF}.

\begin{table}[!htbp]
\centering
\formfactortablestyle

\begin{tabular}{@{}lccc@{}}
\toprule
$F$ & $a$ & $b$ & $M_{\rm pole}$ \,[GeV] \\
\midrule

$f_1$ & 0.007 & $-0.196$ & $m_{K^*}$ \\
$f_2$ & 1.959 & $-1.107$ & $m_{K^*}$ \\
$f_3$ & $-0.028$ & 0.032 & $m_{K_0^*(700)}$ \\
$g_1$ & $-0.458$ & $-0.154$ & $m_{K_1(1270)}$ \\
$g_2$ & 0.021 & $-0.240$ & $m_{K_1(1270)}$ \\
$g_3$ & 3.263 & $-0.270$ & $m_K$ \\

\bottomrule
\end{tabular}

\caption{Form-factor parameters for $\Omega_c^0\to\Xi_c^+$ from Ref.~\cite{Xing:2024okl}.}
\label{tab:OmegacXicFF}

\end{table}



\section{\texorpdfstring{$\Xi^0-\bar\Xi^0$}{Xi oscillations}}
\label{sec:Xiosc}

The possibility of $\Xi^0-\bar\Xi^0$ oscillations was previously considered in Ref.~\cite{McKeen:2015cuz}. Here, following
Refs.~\cite{Beneito:2025ond,BasiBeneito:2026feb}, we briefly derive an indirect constraint from their contribution to $\Lambda -\bar\Lambda$ mixing. Unlike the heavy-baryon transitions considered in the main text, both baryons belong to the light baryon octet, allowing the weak transition to be described within SU(3) B$\chi$PT~\cite{Bijnens:1985kj,Jenkins:1991bt,Borasoy:1998ku}.

The relevant leading-order nonleptonic weak chiral interaction reads~\cite{Bijnens:1985kj,Jenkins:1991bt,Borasoy:1998ku,Ivanov:2021huf,Tandean:2002vy}
\begin{equation}
    \mathcal{L}_{\mathrm{eff}}^{|\Delta S|=1} = a\,\mathrm{Tr}\!\left(\bar {\mathcal B}\{\xi^\dagger h\xi,\mathcal{B}\}\right) + b\,\mathrm{Tr}\!\left(\bar {\mathcal B}[\xi^\dagger h\xi,\mathcal{B}]\right) +\mathrm{h.c.}\,,
\end{equation}
where $\mathcal{B}$ and $\xi$ denote the baryon-octet and Goldstone-boson matrices, respectively, and $h_{23}=1$ is the only nonzero entry of the weak spurion. This interaction contains the term
\begin{equation}
    \mathcal{L}_{\mathrm{eff}}^{|\Delta S|=1} \supset \frac{-a+3b}{\sqrt{6}}\,\bar\Lambda\,\Xi^0 +\mathrm{h.c.}\,.
\label{eq:weakLambdaXi0}
\end{equation}

Two insertions of Eq.~\eqref{eq:weakLambdaXi0}, connected by a $\Xi^0-\bar\Xi^0$ mass insertion, induce, as in Eq.~\eqref{eq:master} with $g^{\rm PC}_{\Xi^0\Lambda}=(-a+3b)/\sqrt6$,
\begin{equation}
    |\delta m_\Lambda^{(\Xi^0)}| \simeq \left| \frac{-a+3b}{\sqrt{6}(m_{\Xi^0}-m_\Lambda)}\right|^2 |\delta m_{\Xi^0}|\,.
\label{eq:deltamLambdaFromXi0}
\end{equation}

Using  $a= 0.56\, G_F\, m_\pi^2\, f =1.68\times 10^{-8}$ GeV and $b= -1.42\,G_F\,m_\pi^2\, f =-4.26\times 10^{-8}$ GeV~\cite{Bijnens:1985kj}, we obtain
\begin{equation}
    |\delta m_\Lambda^{(\Xi^0)}| \simeq 8.8\times10^{-14}\,|\delta m_{\Xi^0}|\,,
\label{eq:Xi0LambdaNumerical}
\end{equation}
where $f=132~{\rm MeV}$ and $m_\pi=m_{\pi^\pm}$.

Using Eq.~\eqref{eq:light_baryon_inputs} and neglecting possible cancellations with other contributions to $\Lambda-\bar\Lambda$ mixing, we obtain the estimated indirect bound
\begin{equation}
    |\delta m_{\Xi^0}|\lesssim10^{-19}\,{\rm GeV}\,.
\label{eq:deltamXi0Constraint}
\end{equation}

The authors of Ref.~\cite{Aitken:2017wie} obtained a stronger estimate, $|\delta m_{\Xi^0}| \lesssim 10^{-22}\, \mathrm{GeV}$, by relating the $|\Delta S|=4$ operator responsible for $\Xi^0-\bar\Xi^0$ oscillations to $|\Delta S|=3$ dinucleon decay through a single Standard Model weak insertion. Their estimate assumes that channels such as $pp\to K^+K^+K^0$, for which no dedicated experimental search exists, are constrained at approximately the same level as the $pp\to K^+ K^+$ channel searched for in Ref.~\cite{Super-Kamiokande:2014hie}.

\bibliographystyle{JHEP}
\bibliography{main}

\end{document}